\documentclass[preprint, showpacs,preprintnumbers,amsmath,amssymb]{revtex4}
\usepackage[dvips]{graphicx}
\usepackage{graphicx}
\usepackage{amsfonts}
\usepackage{bm}
\usepackage{amsmath}
\usepackage{amssymb}
\usepackage{color}
\usepackage[all]{xy}

\def\be{\begin{equation}}
\def\ee{\end{equation}}
\def\bea{\begin{eqnarray}}
\def\eea{\end{eqnarray}}

\begin{document}

\title{Geometrothermodynamics of black holes with a nonlinear source}

\author{Alberto S\'anchez}
\email{asanchez@ciidet.edu.mx} \affiliation{Departamento de
posgrado, CIIDET,\\{\it AP752}, Quer\'etaro, QRO 76000, MEXICO}

\date{\today}

\begin{abstract}

We study thermodynamics and geometrothermodynamics of a particular
black hole configuration with a nonlinear source. We use the
mass as
fundamental equation, from which it follows that the curvature
radius must be considered as a thermodynamic variable, leading to
an extended equilibrium space. Using the formalism of geometrothermodynamics,
 we show that the geometric properties of the thermodynamic
equilibrium space can be used to obtain information about thermodynamic
interaction, critical points and phase transitions. We
show that these results are compatible with the results obtained
from classical black hole thermodynamics.


\end{abstract}


\maketitle

%%%%%%%%%%%%%%%%%%%%%%%%%%%%%%%%%%%%%%%%%%%%%%%%%%%%%%%%%%%%%%%%%%%%%%%%%%%%%%%%%%%%%%%%%%%%%%%%%
%%%%%%%%%%%%%%%%%%%%%%%%%%%%%%%%%% SECTION I %%%%%%%%%%%%%%%%%%%%%%%%%%%%%%%%%%%%%%%%%%%%%%%%%%%
%%%%%%%%%%%%%%%%%%%%%%%%%%%%%%%%%%%%%%%%%%%%%%%%%%%%%%%%%%%%%%%%%%%%%%%%%%%%%%%%%%%%%%%%%%%%%%%%%

\section{Introduction}
\label{intro}

Several regular black hole solutions have been found by coupling
gravity to nonlinear electrodynamics theories. Among various
regular models known to date, especially intriguing are the
solutions to the coupled equations of nonlinear electrodynamics
and general relativity \cite{Ayon2,Bronnikov}. The description of
magnetically charged black hole provides an interesting example of
the system that could be both regular and extremal. In this same
context are the theories of AdS-black holes that consider the
so-called Power Maxwell Invariant (PMI) field
\cite{Hassaine,Hassaine2,Hendi0,Maeda} where an $s$ parameter is
introduced in order to represent a power term in the
electromagnetic action, i.e. $(F_{\mu\nu}F^{\mu\nu})^{s}$, which
reduces to Maxwellian field (linear electromagnetic source) when
$s=1$. Nonlinear electrodynamic theories have been important at
the low-energy limit of heterotic string theory or in the study of
effects loop corrections in quantum electrodynamics \cite{Kats}.

On the other hand, black hole thermodynamics has been the subject
of numerous researches in theoretical physics. The main attraction
lies in the fact that a black hole is the best system to seek the
aspects of quantum gravity and it is expected that the
thermodynamics of black holes help us to know about its
microscopic structure. The norm/gravity correspondence introduced
by Maldacena \cite{Maldacena}, for example, considers the
thermodynamic study of asymptoticaly AdS black holes relating
black holes on the gravity side with the temperature on the field
theory side. Hawking and Page's research \cite{Hawking} showed
that black holes could be assigned entropy through which it is
possible to study  thermodynamic properties such as phase
transitions and interaction; for example, the
Reissner-Nordstr\"om-AdS (RNAdS) black hole in $n+1$ dimensions
\cite{Chamblin1}, where it was found that it has phase
transitions, or AdS black holes, where the cosmological constant
is considered as a new thermodynamic variable
\cite{Kastor}\cite{Dolan1}. However, the full implication of the
gravitational-- thermodynamic connection is not yet apparent
\cite{davies}.

On the other hand, the geometric description of the thermodynamic
properties of black holes has been investigated by using two
different approaches, namely, Weinhold and Ruppeiner's approach
 of thermodynamic geometry \cite{Weinhold,Ruppeiner} and the
mathematical approach called geometrothermodynamics (GTD)
\cite{quevedo2}. Both formalisms use a Riemannian manifold to
define a space of equilibrium states where the thermodynamic
phenomena take place; however, the Weinhold and Ruppeiner
approaches are not Legendre inavariant,  which implies that the
properties of a given thermodynamic system can depend on the
choice of thermodynamic potential. The formalism of GTD, on the
contrary, is a geometric approach which considers the Legendre
invariance and, therefore, describes the properties of the
thermodynamic system independently of the thermodynamic potential,
as in classical thermodynamics. In this work, we will
use GTD in order to study the properties of black holes with a
power Maxwell invariant field.

On the other hand, in black hole thermodynamics the degree of
homogeneity of the fundamental equation is not always considered.
The homogeneity of a thermodynamic system is very important
because it allows us to know if the thermodynamic variables have a
subextensive, extensive or supraextensive character. We show in
this work how the GTD takes into account the homogeneity of the
fundamental equation \cite{Quevedo5}

This work is organized as follows. In Section
\ref{Generalconcepts}, we present the explicit form and review the
main properties of a black hole with a power Maxwell invariant
source, we analyze its fundamental equation and derive its main
thermodynamic properties, considering the cosmological constant as
an additional thermodynamic variable. In Section \ref{GTD}, we
perform a geometrothermodynamic analysis of the corresponding
3-dimensional equilibrium manifold, and show that its
thermodynamic curvature leads to results which are equivalent to
the ones obtained from the analysis of the corresponding heat
capacity. This proves the compatibility between classical black
hole thermodynamics and GTD. Finally, in Section
\ref{conclusions}, we present the conclusions of our work.

%%%%%%%%%%%%%%%%%%%%%%%%%%%%%%%%%%%%%%%%%%%%%%%%%%%%%%%%%%%%%%%%%%%%%%%%%%%%%%%%%%%%%%%%%%%%%%%%%
%%%%%%%%%%%%%%%%%%%%%%%%%%%%%%%%%% SECTION II %%%%%%%%%%%%%%%%%%%%%%%%%%%%%%%%%%%%%%%%%%%%%%%%%%%
%%%%%%%%%%%%%%%%%%%%%%%%%%%%%%%%%%%%%%%%%%%%%%%%%%%%%%%%%%%%%%%%%%%%%%%%%%%%%%%%%%%%%%%%%%%%%%%%%

\section{Black holes with nolinear
sources and theirs thermodynamics} \label{Generalconcepts}

The general action that describes Einstein-PMI gravity is given by
the expression \cite{Hassaine2},

\bea \label{action} I=-\frac{1}{16\pi}\int_M{d^{n+1}x
\sqrt{-g}\Big[R+\frac{n(n-1)}{l^2}+\Big(F_{\mu \nu} F^{\mu
\nu}\Big)^s \Big]\,,}\eea where $F_{\mu \nu}$ represents the
electromagnetic field tensor, $l$ is related to the  cosmological
constant by the expression $\Lambda=-\frac{3}{l^2}$ and $s$ is
called nonlinearity parameter.

Considering a spherically symmetric spacetime with line element,

\bea \label{metric} ds^2=-f(r)dt^2+\frac{dr}{f(r)}+r^2
d\Omega^2{}_{d-2}\,,\eea where $d\Omega^2{}_{d-2}$ stands for the
standard element on $S^d$. Using the equations obtained from
the variation of the action (\ref{action}), we get the next
solution for a black hole with PMI source \cite{Hassaine,Hendi}

\bea f(r)=1+\frac{r^2}{l^2}-\frac{m}{r{}^{n-2}}
+\frac{(2s-1)^2\Big[\frac{(n-1)(2s-n)^2 q^2}{(n-2)(2s-1)^2}
\Big]^s}{(n-1)(n-2s) r{}^{\frac{2(ns-3s+1)}{2s-1}}}. \eea

Here, $m$ and $q$ are related to the ADM mass $M$ and the electric
charge $Q$ by means of the relation,

\bea \label{equ31} m&=&\frac{16\pi M }{(n-1)\omega_{n-1}}\,,\\
\label{equ312} q&=&\Bigg[\frac{8\pi}{\sqrt{2} s
\omega_{n-1}}\Bigg]^{\frac{1}{2s-1}}\Bigg[\frac{n-2}{n-1}
\Bigg]^{\frac{1}{2}}\frac{(2s-1)^{\frac{2s-2}{2s-1}}}{n-2s}Q^{\frac{1}{2s-1}}.\eea

The roots of the lapse function $f (r)$ ($g_{tt} = 0$) define the
horizons $r = r_{\pm}$ of the spacetime. In particular, the null
hypersurface $r = r_+$ can be shown to correspond to an event
horizon, which in this case is also a Killing horizon, whereas the
inner horizon at $r_{-}$ is a Cauchy horizon. Therefore, from
$f(r_+) = 0$ we get the black hole mass, $M$,

\bea \label{mass}M(r_+,Q)=\frac{(n-1)\omega_{n-1}
}{16\pi}\Bigg[r_+{}^{n-2}+\frac{r_+{}^n}{l^2}-f_nr_+{}^{\frac{(2s-n)}{2s-1}}Q^{\frac{2s}{2s-1}}\Bigg]\,,
\eea with

\bea \label{factor}
f_n=\frac{(2s-1)^{2-2s}(n-1)^{s-1}(2s-n)^{2s-1}}{(n-2)^s}\Bigg[\frac{8\pi}{\sqrt{2}
s \omega_{n-1}}\Bigg]^{\frac{2s}{2s-1}}\Bigg[\frac{n-2}{n-1}
\Bigg]^{s}\frac{(2s-1)^{\frac{2s(2s-2)}{2s-1}}}{(n-2s)^{2s}}\,.
\eea

From the area-entropy relationship, $S=\frac{\omega_{n-1}
r_+{}^{n-1}}{4}$, equation (\ref{mass}) can be rewritten as

\bea \label{mass2}
M(S,Q)=\frac{[4]^{\frac{n}{n-1}}(n-1)}{\pi}\omega_{n-1}^{\frac{1}{n-1}}\Bigg[S^{\frac{n-2}{n-1}}+S^{\frac{n}{n-1}}l^{-2}-\tilde{f}_n
S^{\frac{2s-n}{(n-1)(2s-1)}}Q^{\frac{2s}{2s-1}}\Bigg],\eea where,

\bea \label{factor2}
\tilde{f}_n=\Big[\frac{4}{\omega_{n-1}}\Big]^{\frac{6s-2sn-2}{(2s-1)(n-1)}}f_n\,,
\eea with $\omega_{n-1}=(2\pi^{n/2})/\Gamma(n/2)$. In order to
avoid inconsistent results, we will assume that $s>\frac{n}{2}$.

The equation (\ref{mass2}) is the fundamental equation for black
holes with PMI source. This equation relates all the thermodynamic
variables entering the black hole metric. In order to correctly
describe the thermodynamic properties of black holes, it is
necessary to impose the condition that the corresponding
fundamental equation be a quasi-homogeneous function
\cite{Quevedo5,Quevedo6}. As we can observe, the fundamental
equation (\ref{mass2}) is an inhomogeneous function in the
extensive variables $S$ and $Q$, e.i., the rescaling $S
\rightarrow S  \lambda^{\beta_S}$ and $Q \rightarrow Q
\lambda^{\beta_Q}$ , where $\lambda$ and the $\beta$'s are real
constants, does not fulfill the condition$ M(\lambda^{\beta_S}
S,\lambda^{\beta_Q} Q ) = \lambda^{\beta_M} M(S, Q)$. However, if
we consider $l$ also as a thermodynamic variable, which rescales
as $ l \rightarrow \lambda^{\beta_l} l$, the quasi-homogeneity
condition $M(\lambda^{\beta_S} S,\lambda^{\beta_Q} Q,
\lambda^{\beta_l} l ) = \lambda^{\beta_M} M(S, Q, l)$ holds if the
relationships,

\bea \label{hconditions} \beta_S=\frac{n-1}{n-2}\,,\quad \quad
\beta_Q=\frac{(2s-1)(n^2-2n-s+1)}{s(n-2)(2n-1)}\,,\quad \quad
\beta_l=\frac{1}{n-2}\quad \quad \beta_M=1\,,\eea are fulfilled.

The physical parameters of the black hole with PMI source satisfy
the first law of black hole thermodynamics \cite{davies},

\bea \label{flaw} dM=TdS+\Phi dQ+L dl\,,\eea where $T$ is the
Hawking temperature, which is proportional to the surface gravity
on the horizon, $\Phi$ is the electric potential and $L$ is the
thermodynamic variable dual to $l$. The thermodynamic equilibrium
conditions are given by the expressions,

\bea \label{c2} T=\frac{\partial M}{\partial S}\,,\quad  \quad
\Phi=\frac{\partial M}{\partial Q}\,,\quad \quad L=\frac{\partial
M}{\partial l}\,.\eea These expressions allow us to compute the
explicit form of the corresponding intensive variables:

\bea \label{temperature} T&=&\Omega_n
\Bigg[\frac{n-2}{n-1}S^{-\frac{1}{n-1}}+\frac{n}{n-1}\frac{S^{\frac{1}{n-1}}}{l^2}-\frac{(2s-n)\tilde{f}_n}{(n-1)(2s-1)}S^{\frac{4s-2sn-1}{(n-1)(2s-1)}}Q^{\frac{2s}{2s-1}}
\Bigg]\,,\\ \label{potential} \Phi &=&-\frac{2s\Omega_n}{ (2s-1)}
\tilde{f}_n S^{\frac{2s-n}{(n-1)(2s-1)}}Q^{\frac{1}{2s-1}}\,, \\
\label{duall} L&=&-\frac{2\Omega_n}{l^3}
S^{\frac{n}{(n-1)(2s-1)}}\,,\eea with
$\Omega_n=\frac{[4]^{\frac{n}{n-1}}(n-1)}{\pi}\omega_{n-1}^{\frac{1}{n-1}}$.
The behavior of the temperature $T$, electrical potential $\Phi$
and the thermodynamic variable $L$  in terms of the entropy $S$ is
shown in Figures 1, 2, 3 y 4 for fixed values of the charge and
the variable $l$.

\begin{figure}[h]\begin{center}{\includegraphics[scale=0.25]{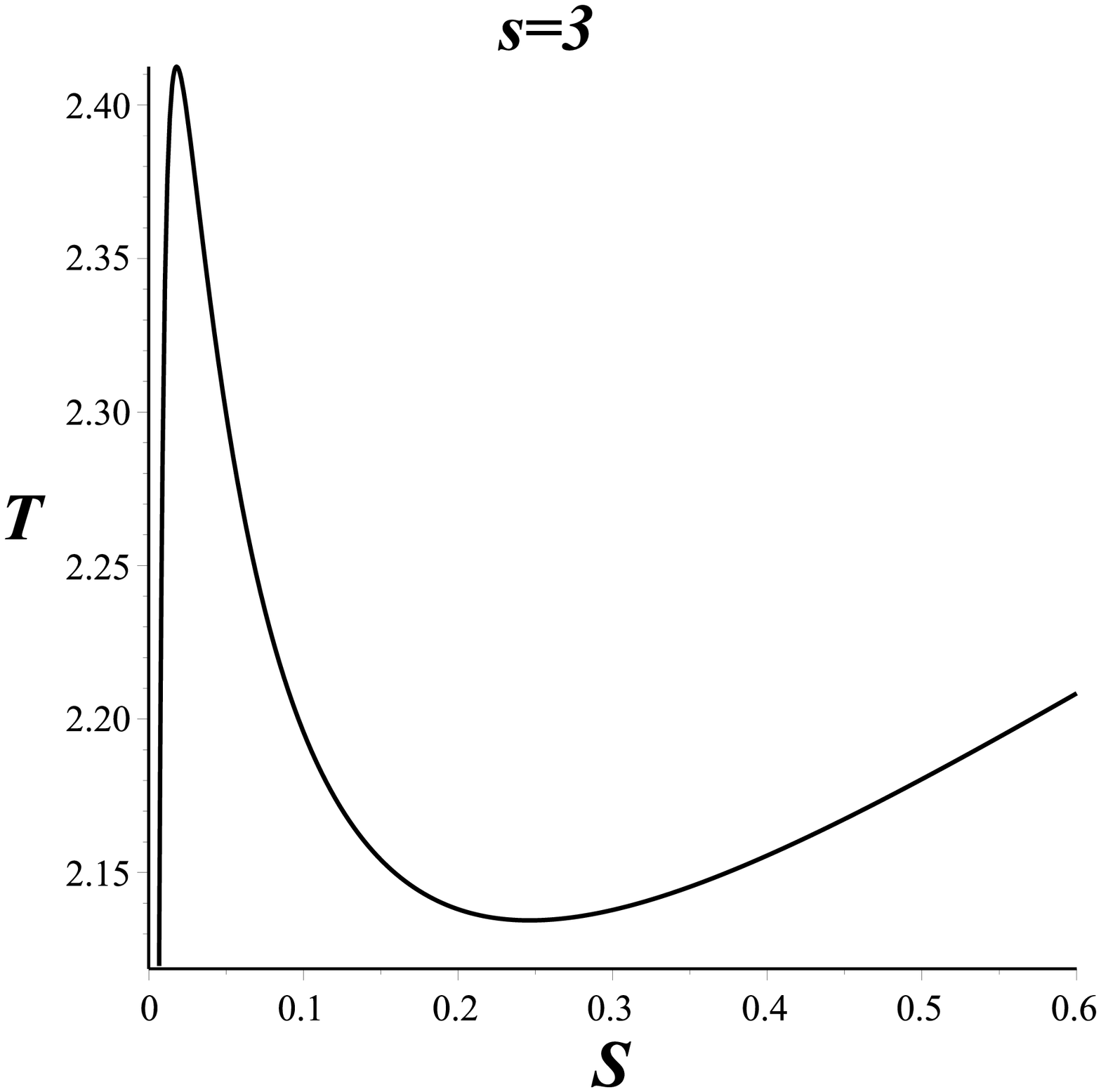}
\includegraphics[scale=0.25]{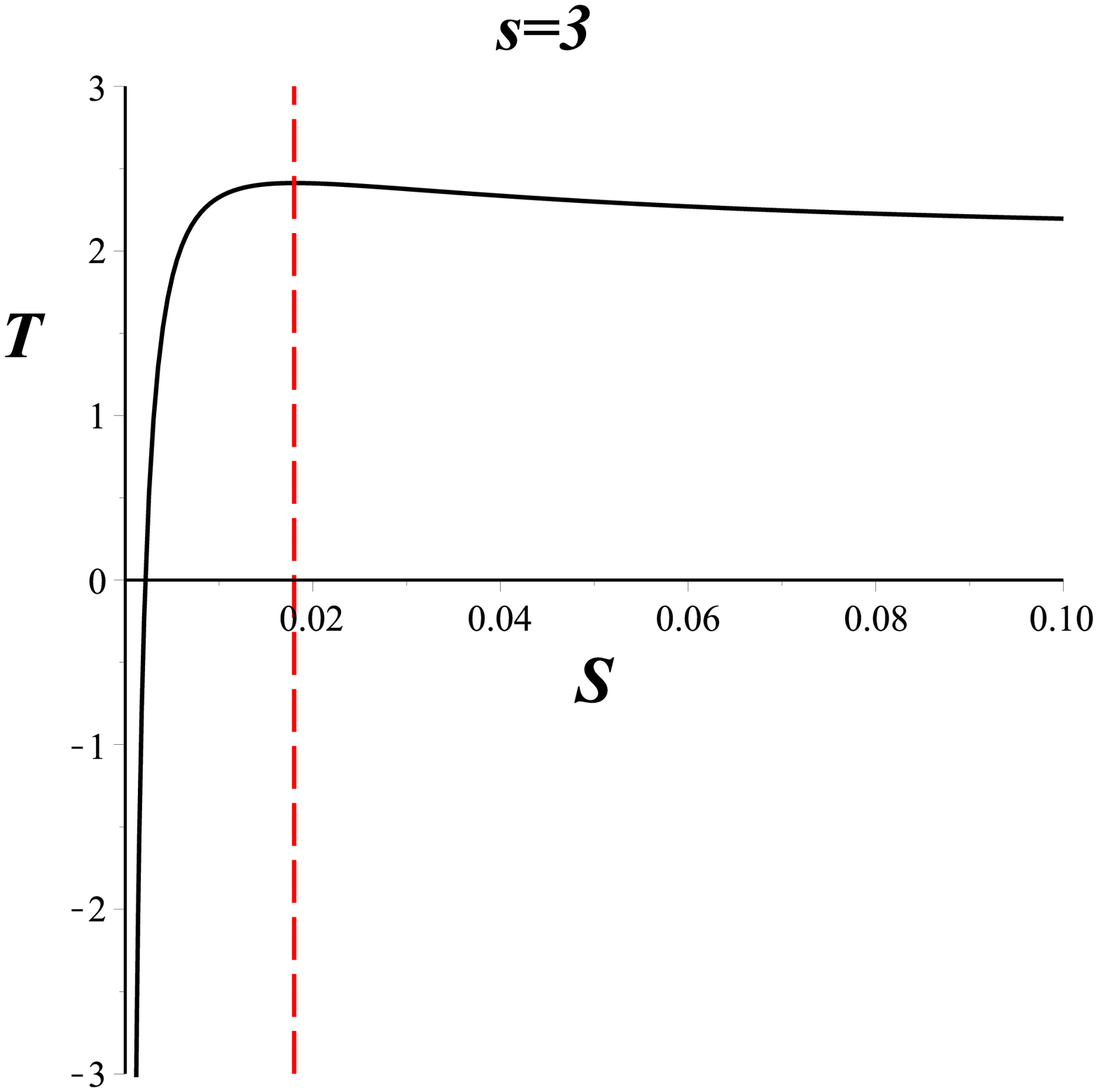}\includegraphics[scale=0.25]{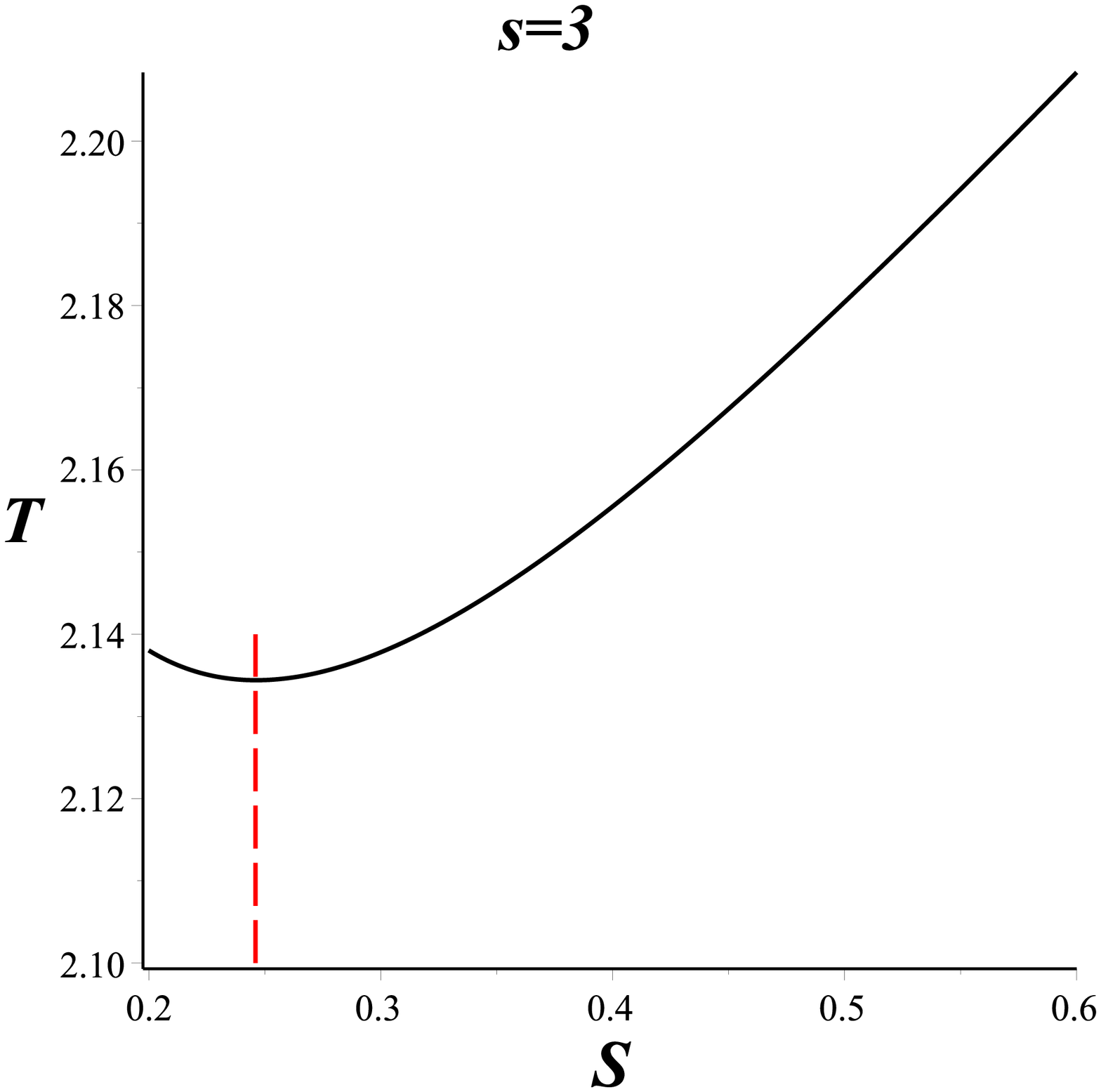}
\caption{{\small Temperature $T$ as a function of entropy $S$,
where we have considered $n=4$, $l=Q=1$ and $s=3$.}}}
\end{center}\end{figure}

\begin{figure}[h]\begin{center}{\includegraphics[scale=0.25]{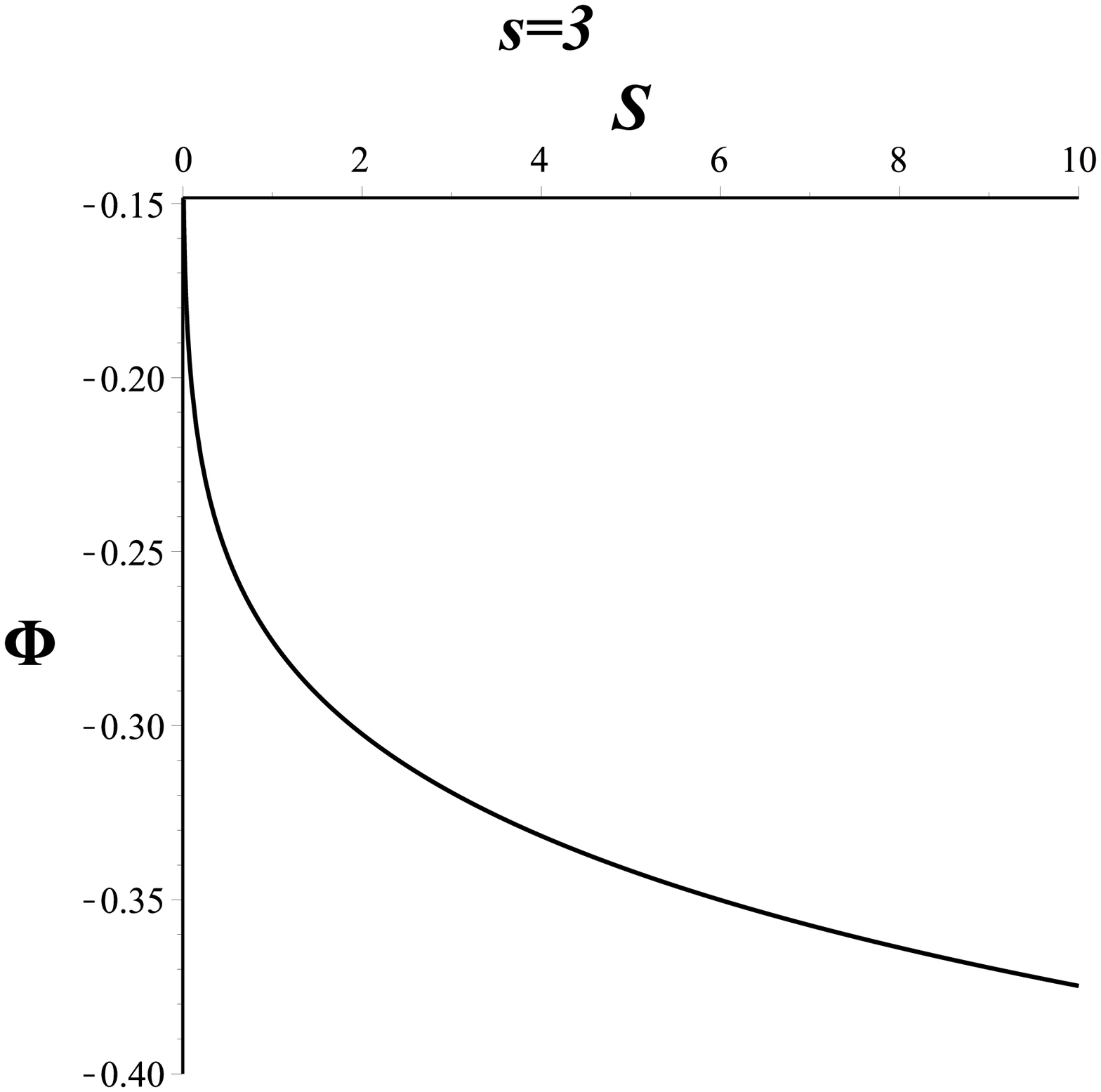}\includegraphics[scale=0.25]{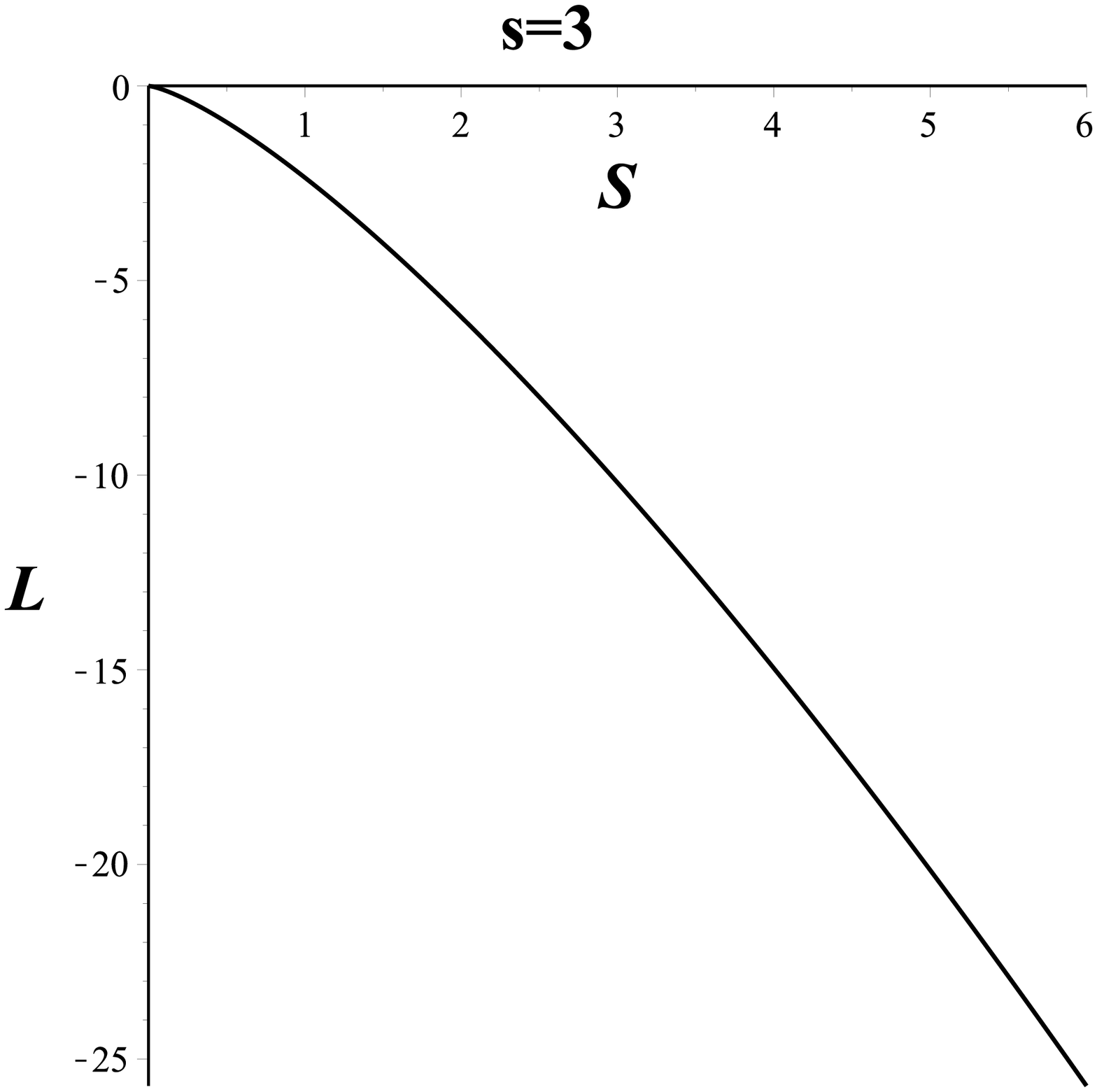}
\caption{{\small Electrical Potential (left) and  the
thermodynamic variable $L$ (right) as a function of entropy $S$,
where we have considered $n=4$, $l=Q=1$ and $s=3$.}}}
\end{center}\end{figure}

\begin{figure}[h]\begin{center}{\includegraphics[scale=0.25]{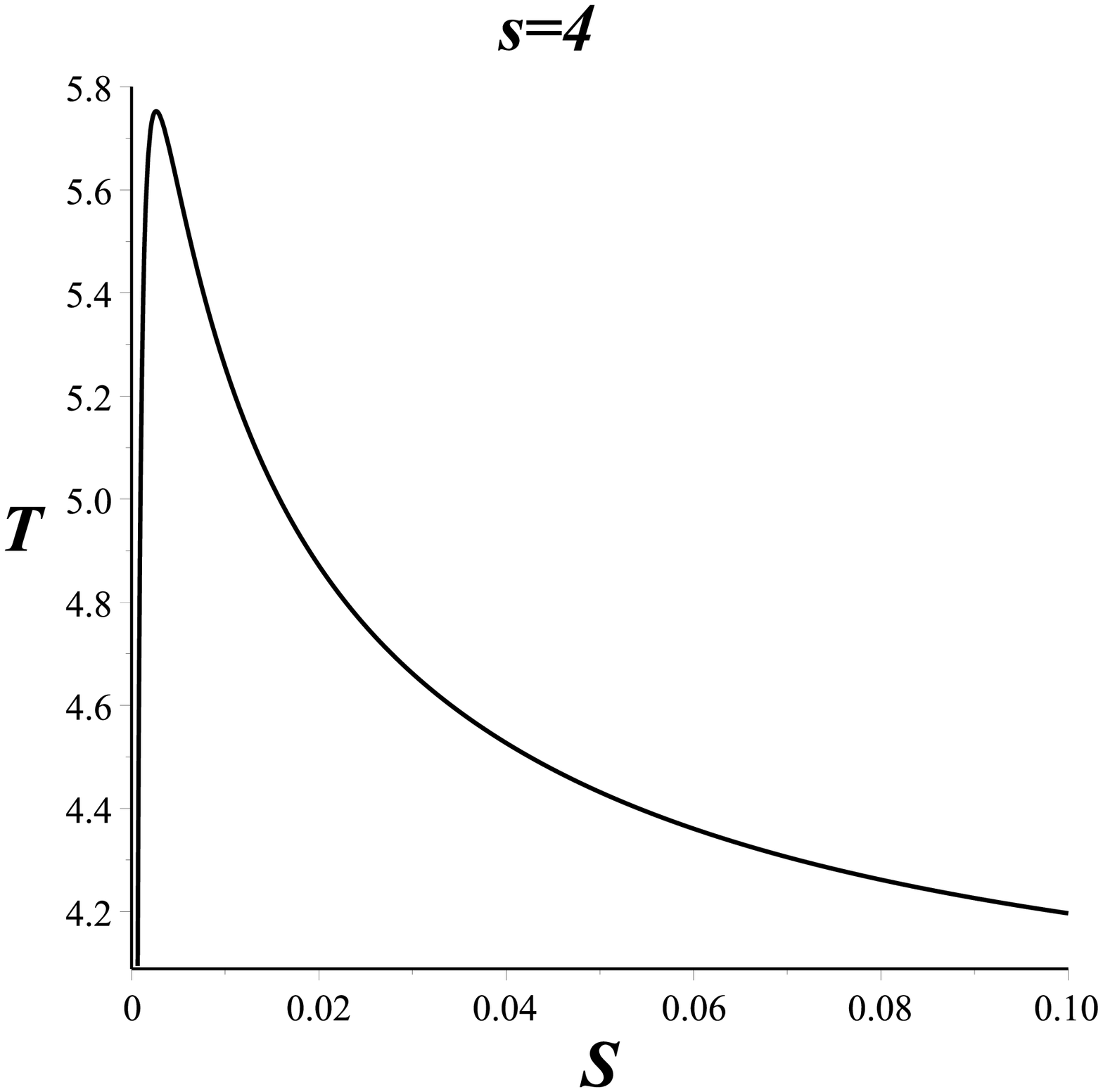}
\includegraphics[scale=0.25]{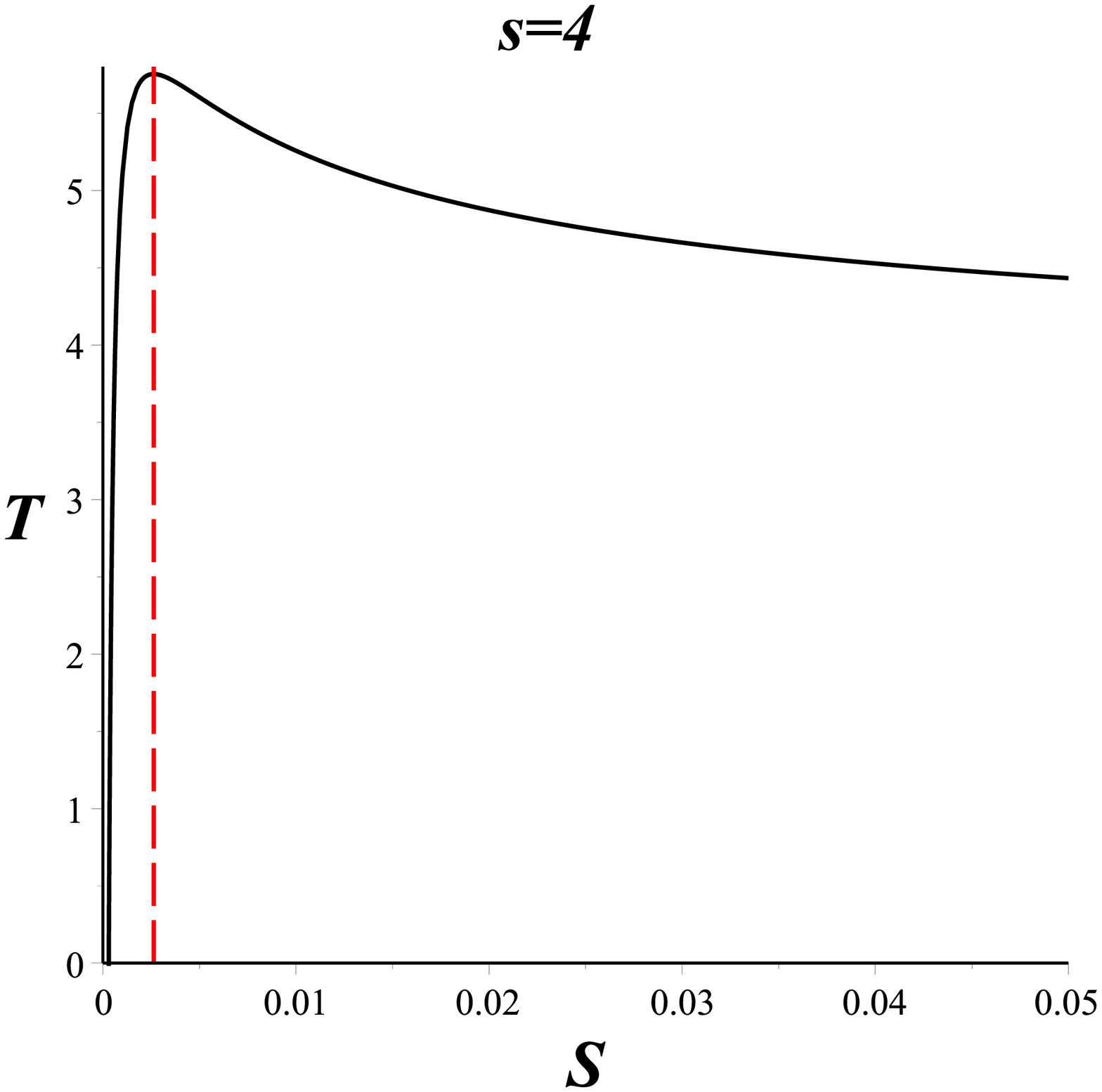}\includegraphics[scale=0.25]{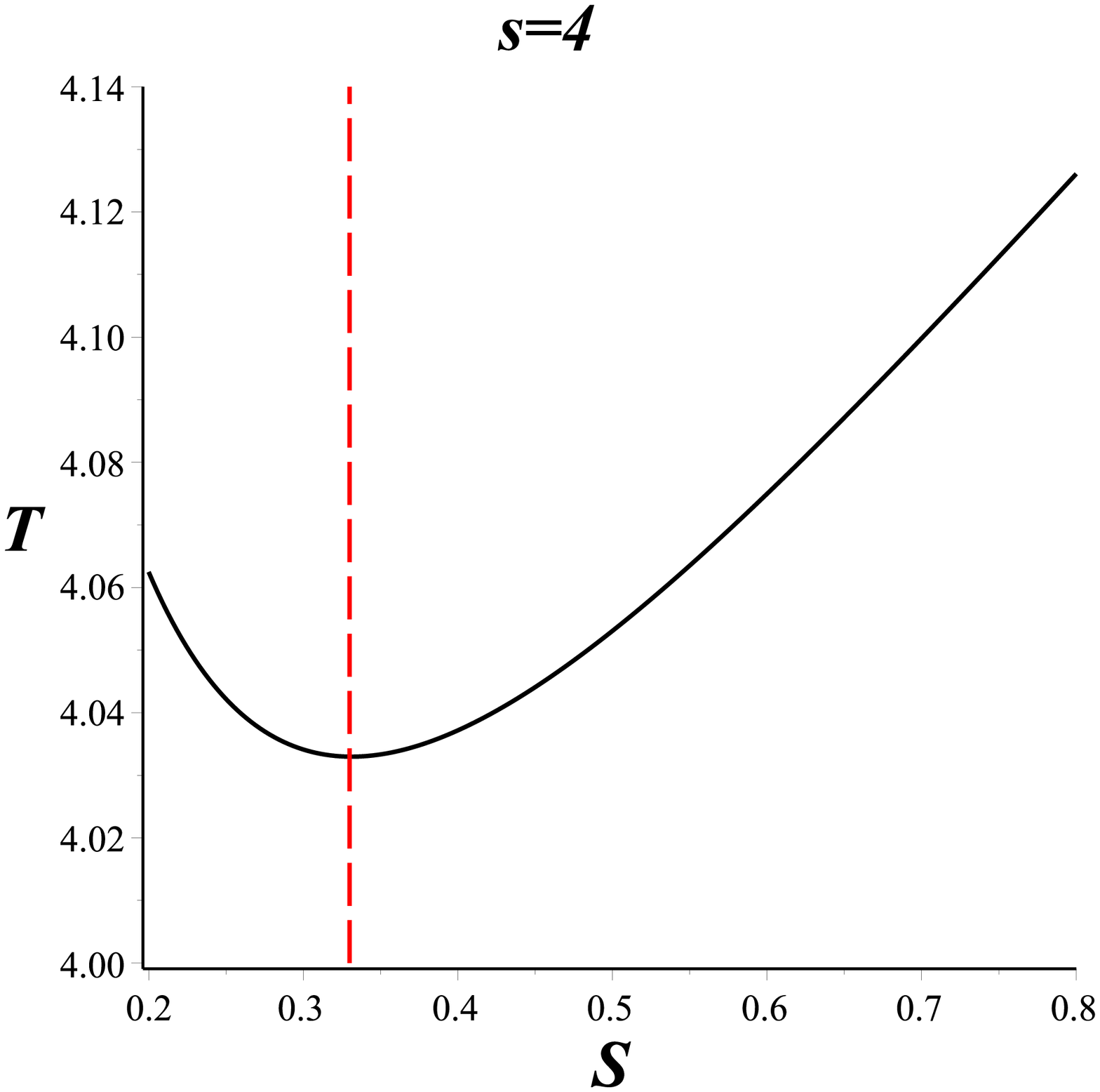}
\caption{{\small Temperature $T$ as a function of entropy $S$,
where we have considered $n=5$, $l=Q=1$ and $s=4$.}}}
\end{center}\end{figure}

\begin{figure}[h]\begin{center}{\includegraphics[scale=0.25]{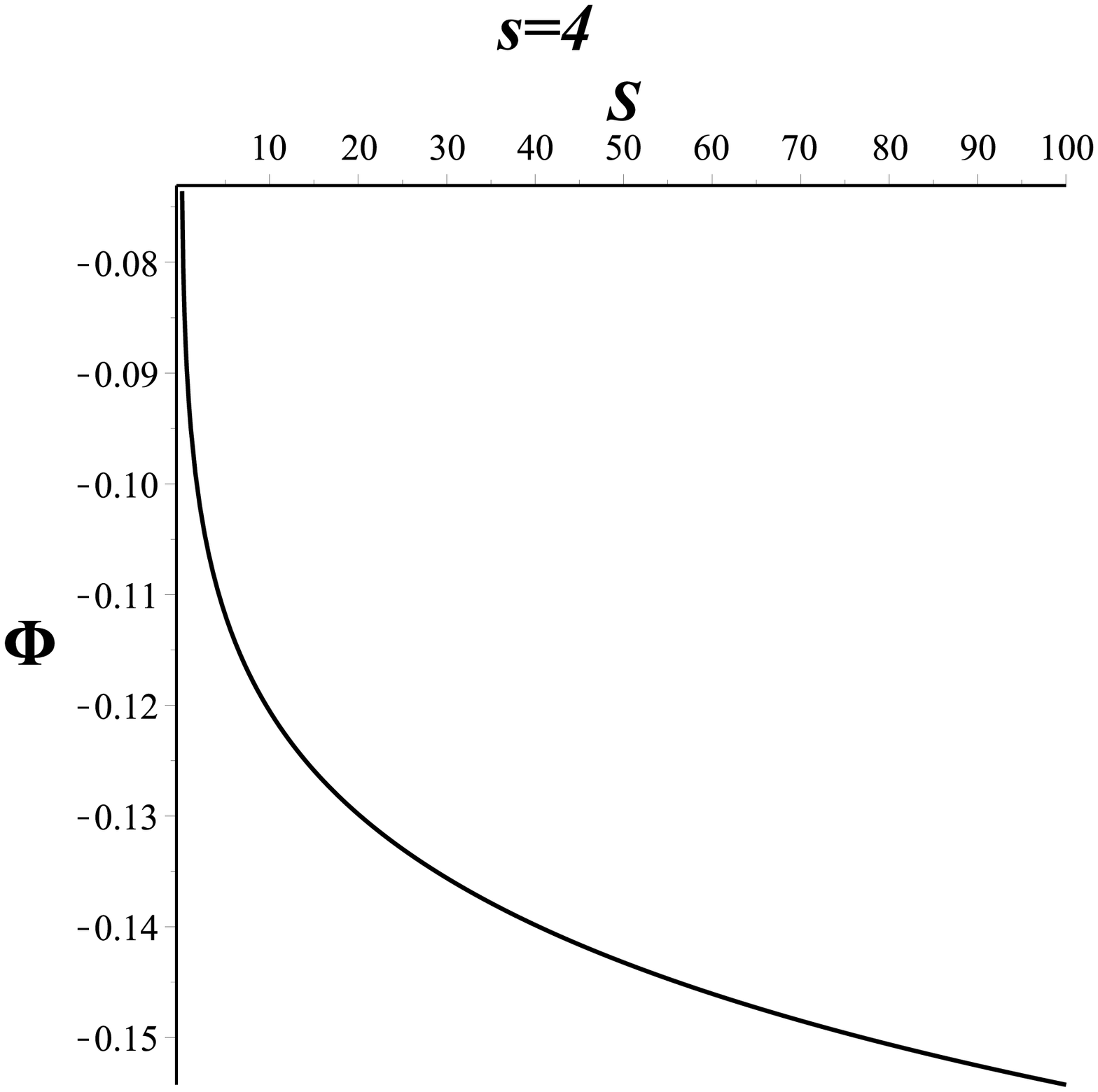}\includegraphics[scale=0.25]{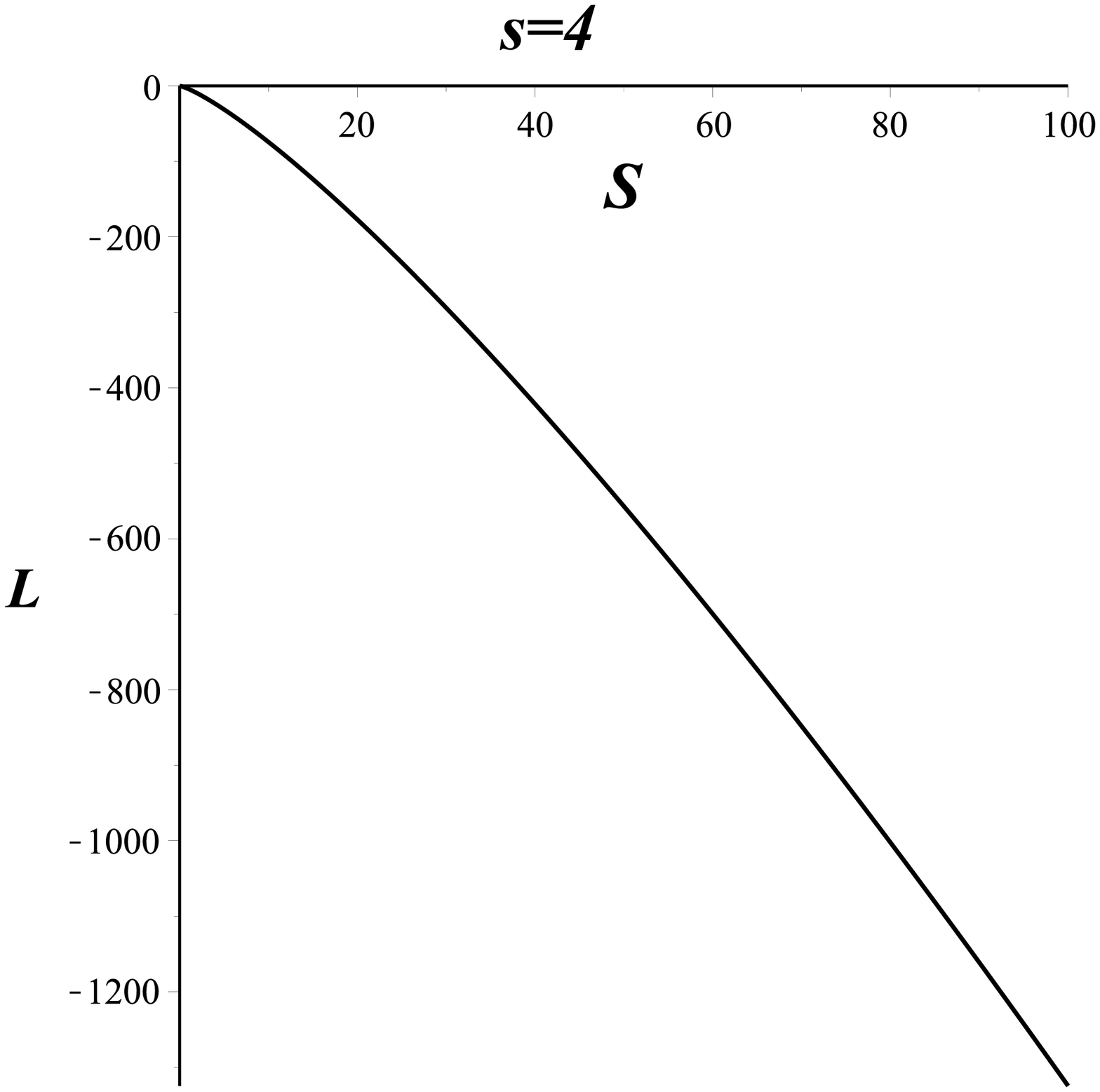}
\caption{{\small Electrical Potential (left) and  the
thermodynamic variable $L$ (right) as a function of entropy $S$,
where we have considered $n=5$, $l=Q=1$ and $s=4$.}}}
\end{center}\end{figure}

It is easy to show that the temperature (\ref{temperature} )
coincides with the Hawking temperature. The temperature increases
rapidly as a function of the entropy $S$ until it reaches its
maximum value. Then, as the entropy increases, the temperature
becomes a monotonically decreasing function until it reaches
another point from which the temperature increases again. The heat
capacity at constant values of $Q$ and  $l$ is computed by the
expression,

\bea \label{c3} C_{{}_{Q\,,l}}=T\Big(\frac{\partial S}{\partial T}
\Big)_{Q\,,l}=\frac{\Big(\frac{\partial M}{\partial S} \Big)_{Q\,,
l}}{\Big(\frac{\partial^2 M}{\partial S^2} \Big)_{Q\,,l}}\,,\eea
where the subscript indicates that derivatives are calculated
keeping $Q$ and $l$ constants. The heat capacity that corresponds
to the fundamental equation (\ref{mass2}) is given by the
expression

\bea \label{c33} C_{{}_{Q\,,l}}=\frac{(n-1)\Big[l^2 (2s-n)
S^{\frac{4s-2ns-1}{(n-1)(2s-1)}} Q^{\frac{2s}{2s-1}}\tilde{f}_n
-(2s-1)[(n-2) l^2+n] S^{-\frac{1}{(n-1)}}\Big]}{(2s-1)\Big[l^2
(n-2) S^{-\frac{n}{(n-1)}}-n S^{-\frac{(n-2)}{(n-1)}}+
\frac{(2s-n)(4s-2sn-1)}{(2s-1)^2}l^2\tilde{f}_n
S^{\frac{6s-4ns+n-2}{(n-1)(2s-1)}}Q^{\frac{2s}{2s-1}}\Big]}\,,\eea

According to Ehrenfest's classification \cite{callen}, second
order phase transitions occur at those points where the heat
capacity diverges, i.e., for

\bea \label{divergeC}  l^2 (n-2) S^{-\frac{n}{(n-1)}}-n
S^{-\frac{(n-2)}{(n-1)}}+
\frac{(2s-n)(4s-2sn-1)}{(2s-1)^2}l^2\tilde{f}_n
S^{\frac{6s-4ns+n-2}{(n-1)(2s-1)}}Q^{\frac{2s}{2s-1}}=0\,.\eea

This behavior is depicted in figure 5

\begin{figure}[h]\begin{center}{\includegraphics[scale=0.25]{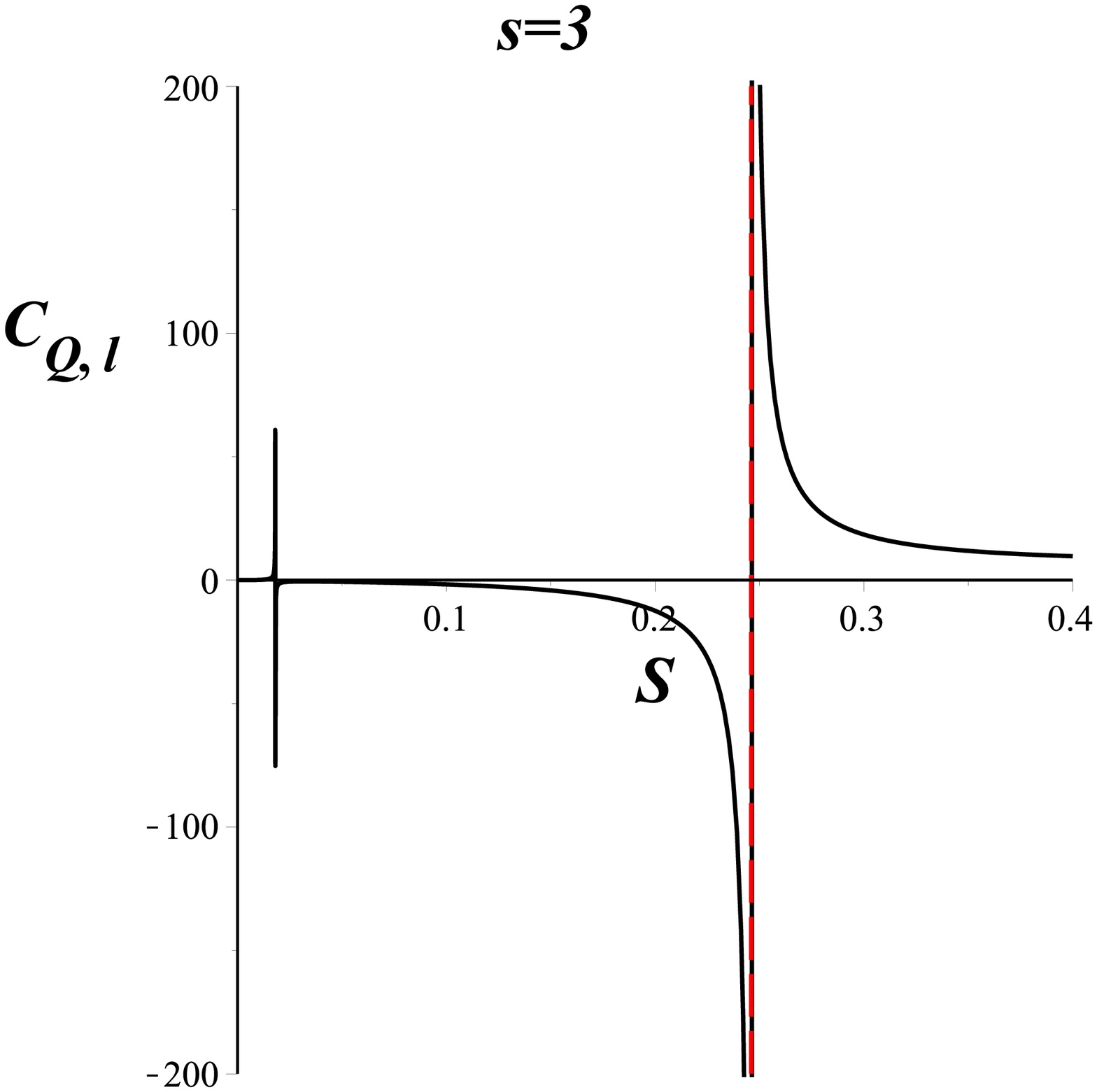}
\includegraphics[scale=0.25]{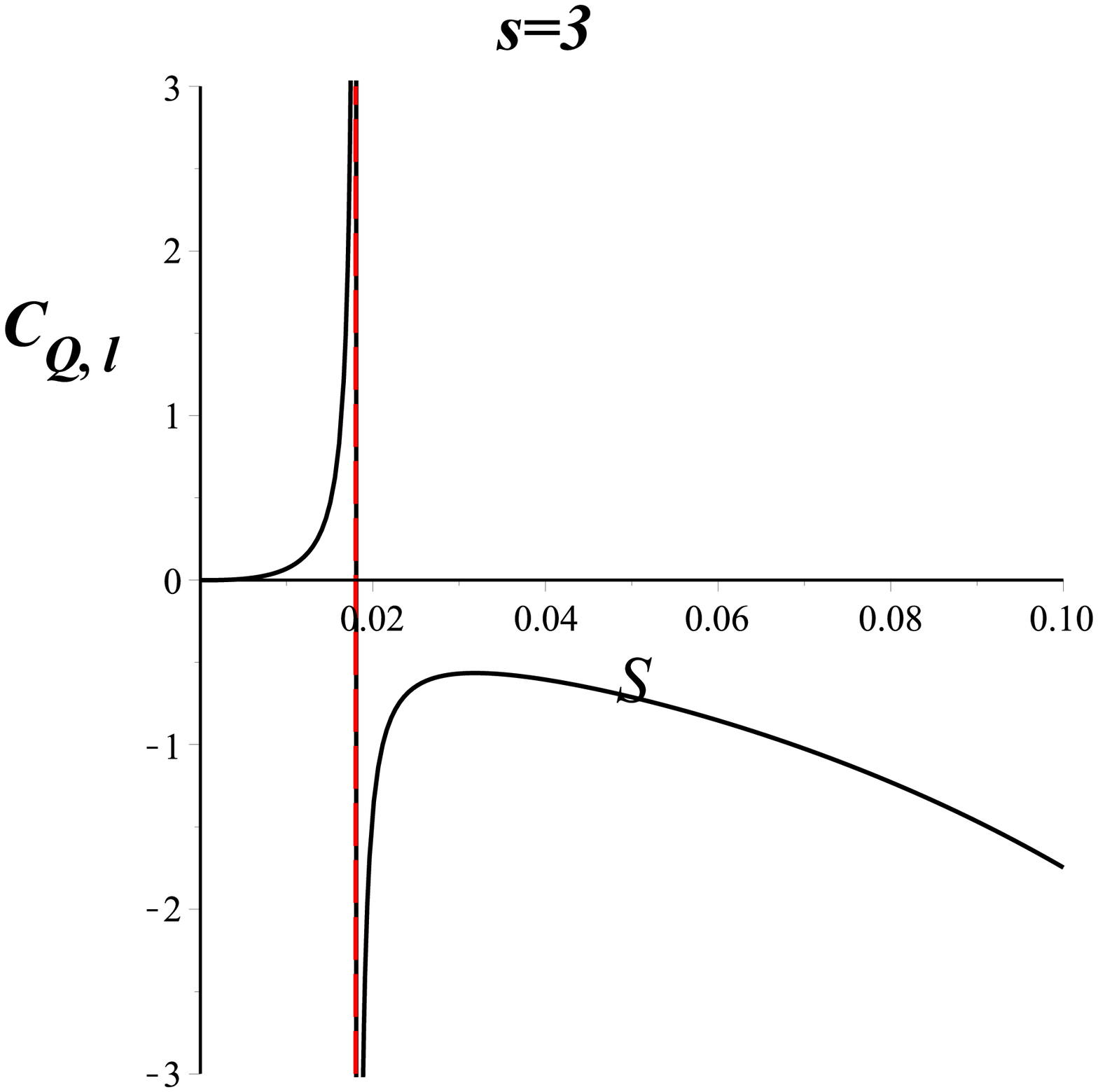}\includegraphics[scale=0.25]{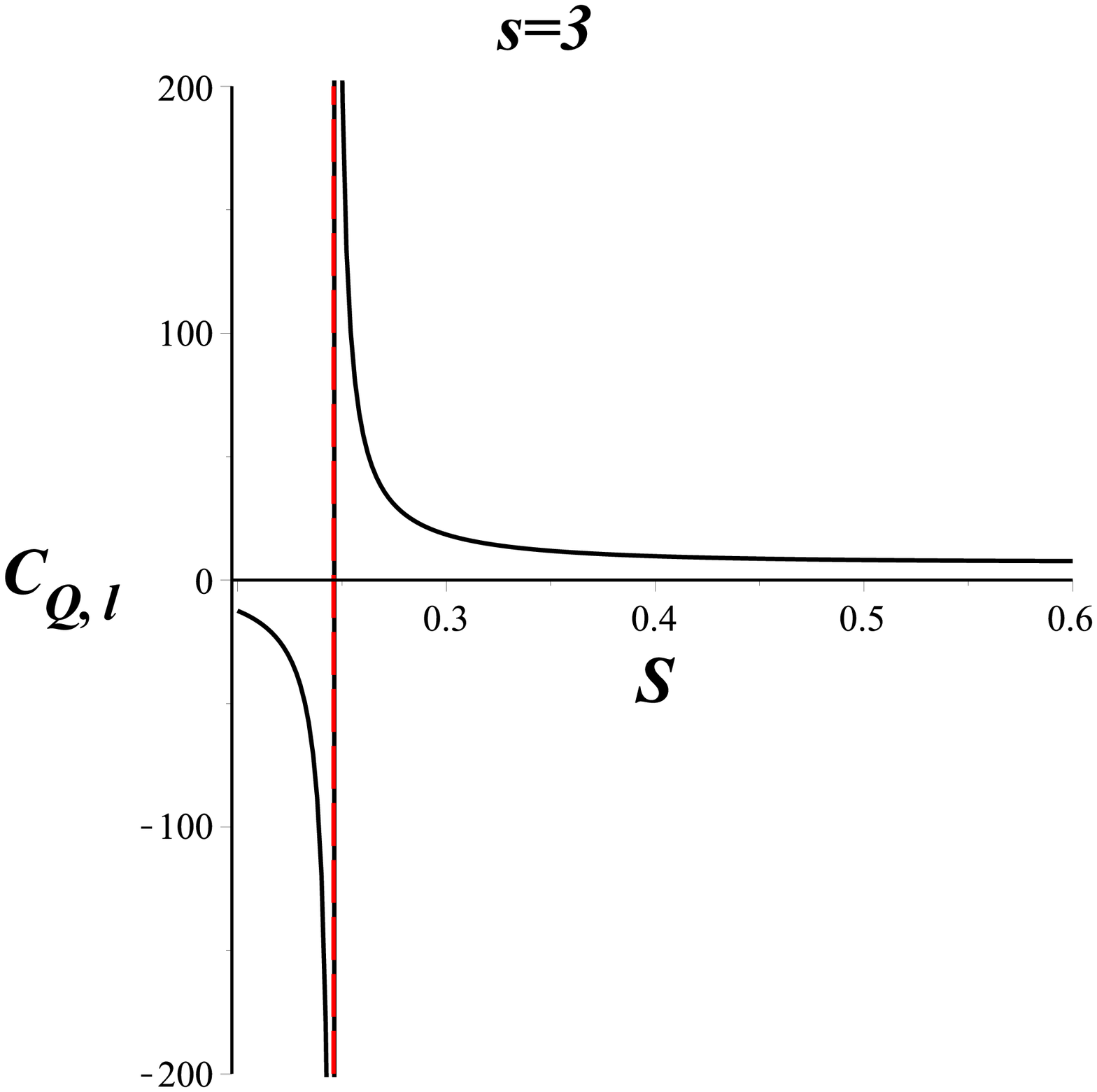}
\caption{{\small Heat Capacity $C_{Q\,,l}$ as a function of
entropy $S$, where we have considered $n=4$, $l=Q=1$ and $s=3$.}}}
\end{center}\end{figure}

\begin{figure}[h]\begin{center}{\includegraphics[scale=0.25]{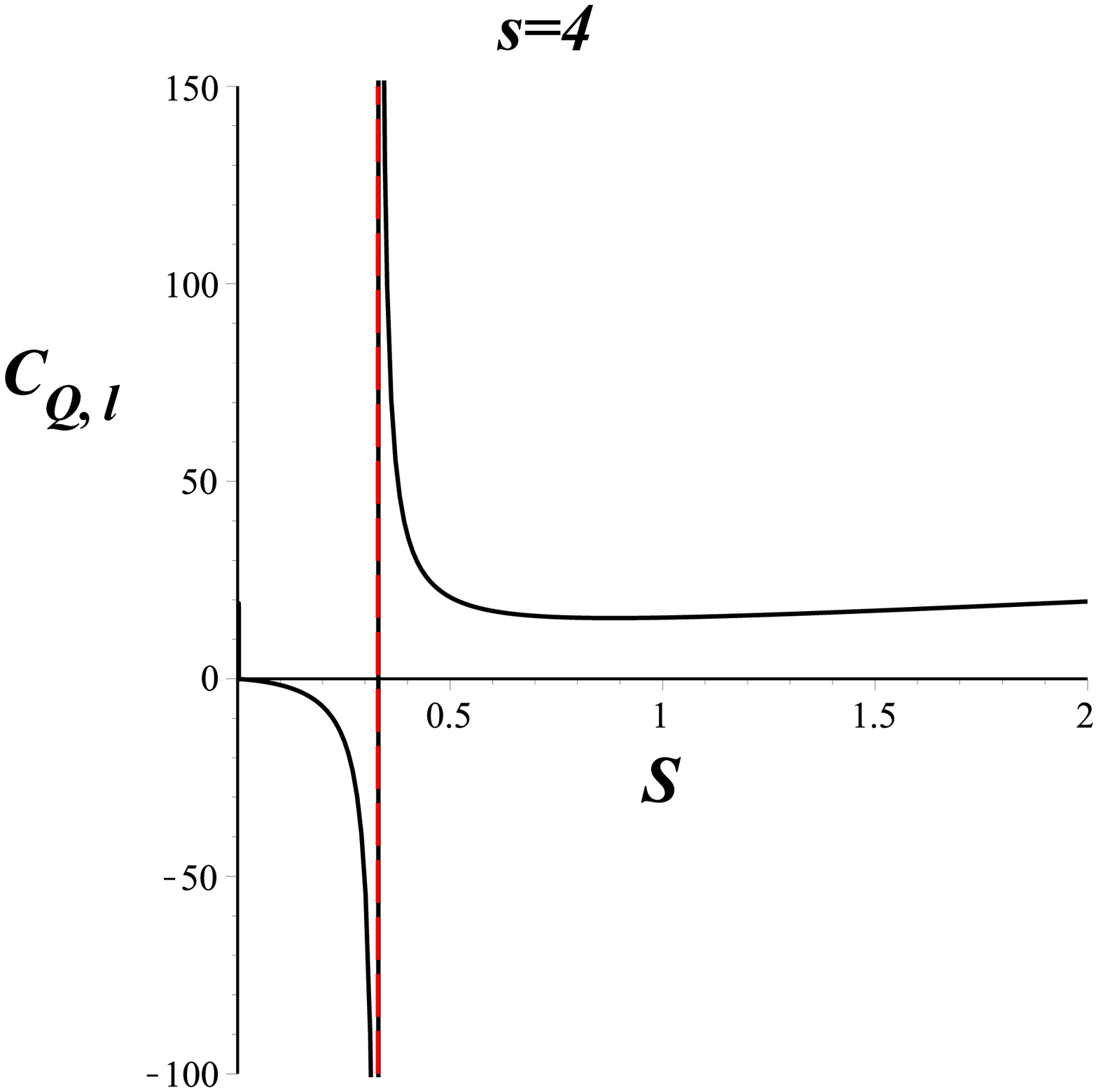}
\includegraphics[scale=0.25]{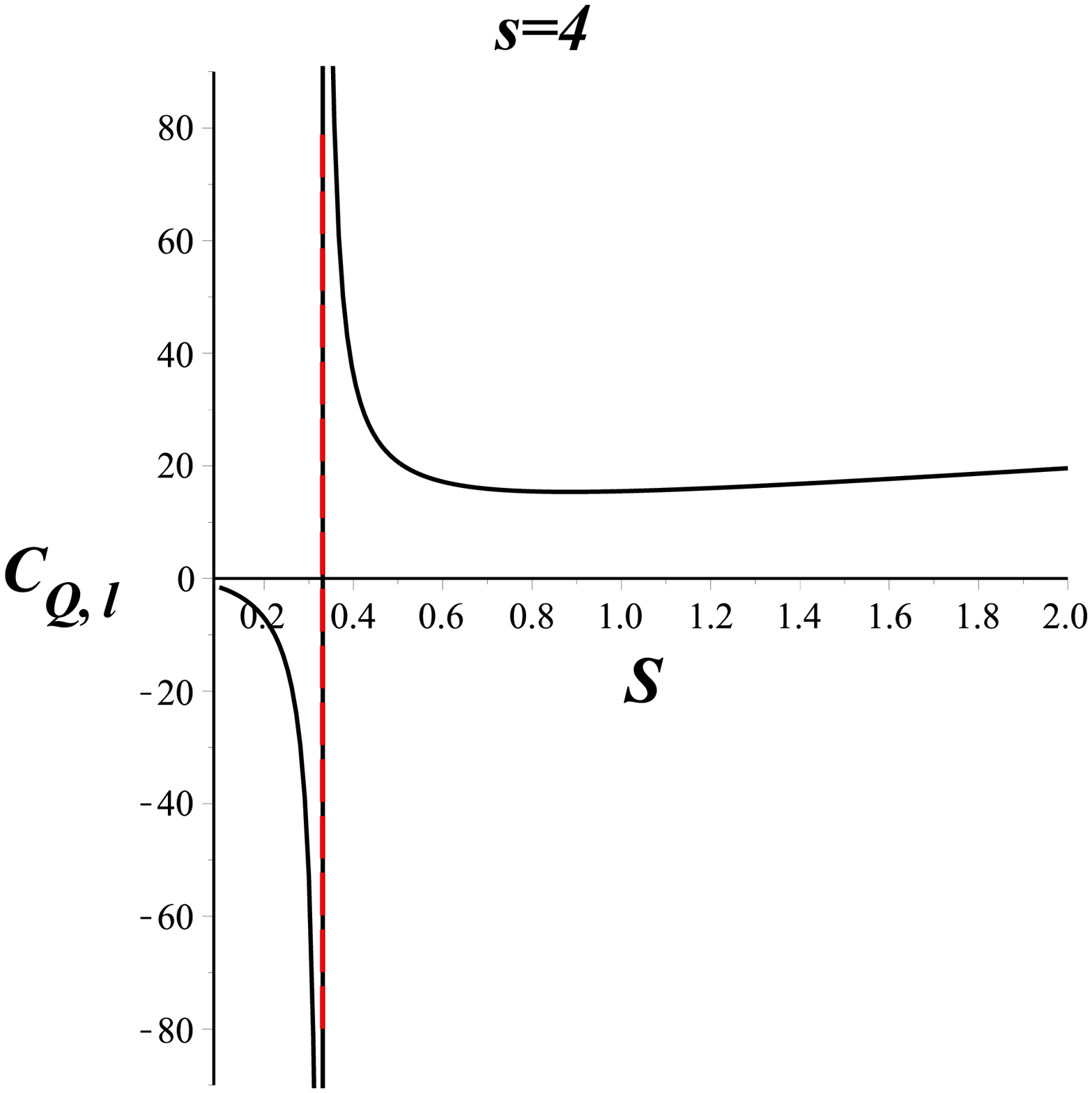}\includegraphics[scale=0.25]{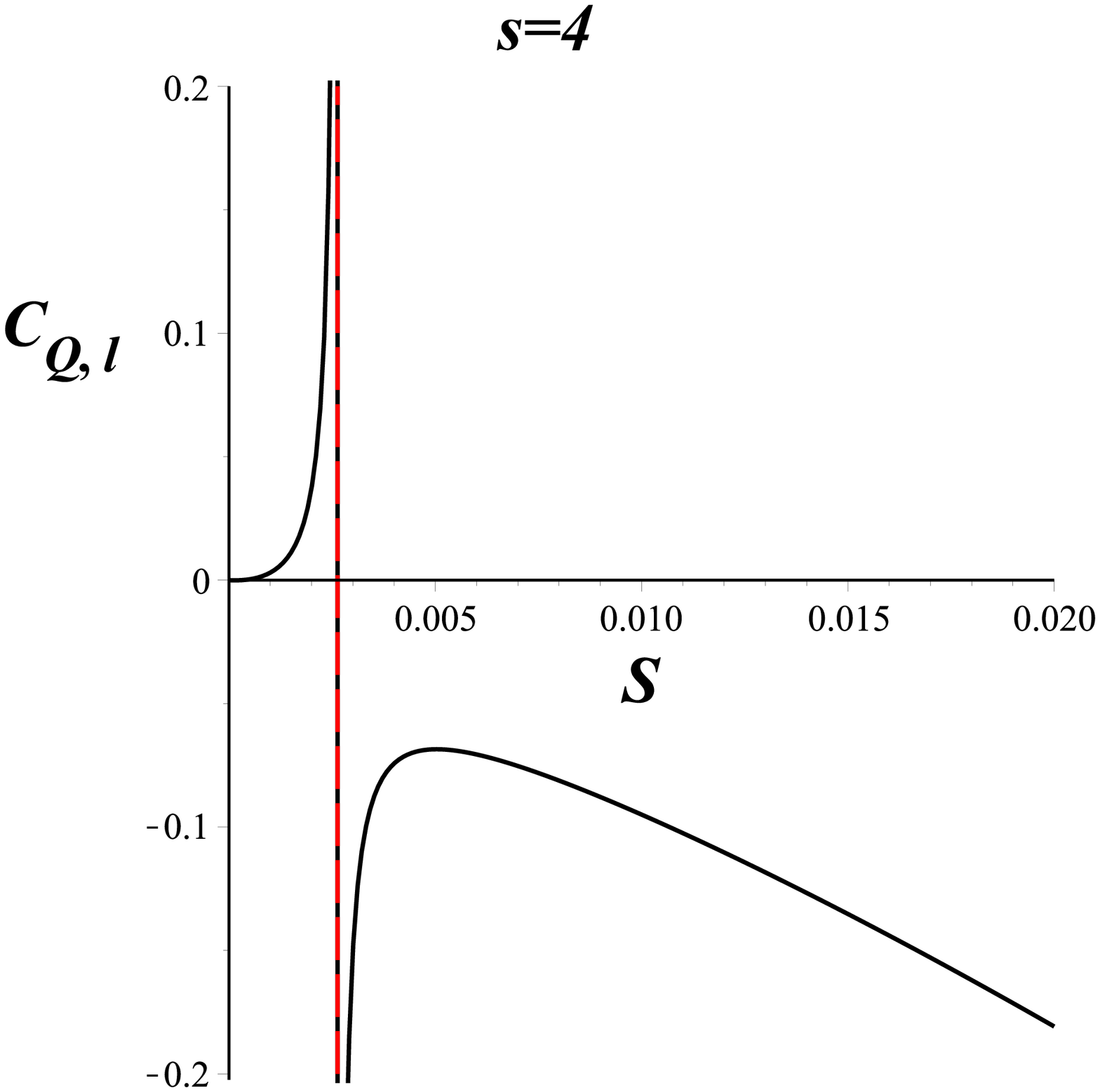}
\caption{{\small Heat Capacity $C_{Q\,,l}$ as a function of
entropy $S$, where we have considered $n=5$, $l=Q=1$ and $s=4$.}}}
\end{center}\end{figure}

We can see from the graphics 1,3,5 and 7 that in the region with
positive temperature the heat capacity is positive, indicating
that the black hole is stable in this region. At the maximum
(minimum) value of the temperature, the heat capacity diverges and
changes spontaneously its sign from positive to negative. This
indicates the presence of second order phase transitions, which
are accompanied by a transition into a region of instability.

\begin{figure}[h]\begin{center}{\includegraphics[scale=0.25]{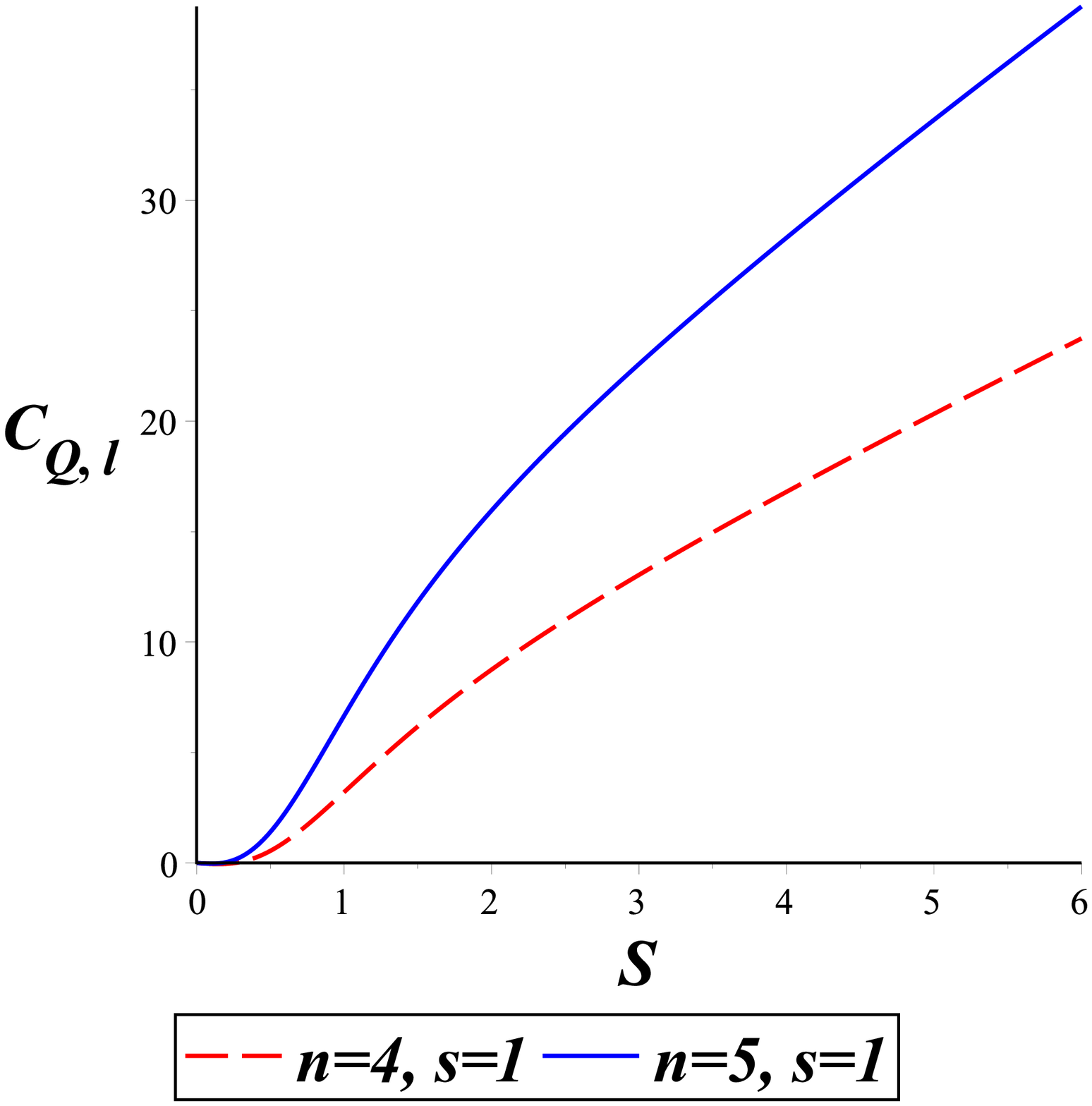}
\includegraphics[scale=0.25]{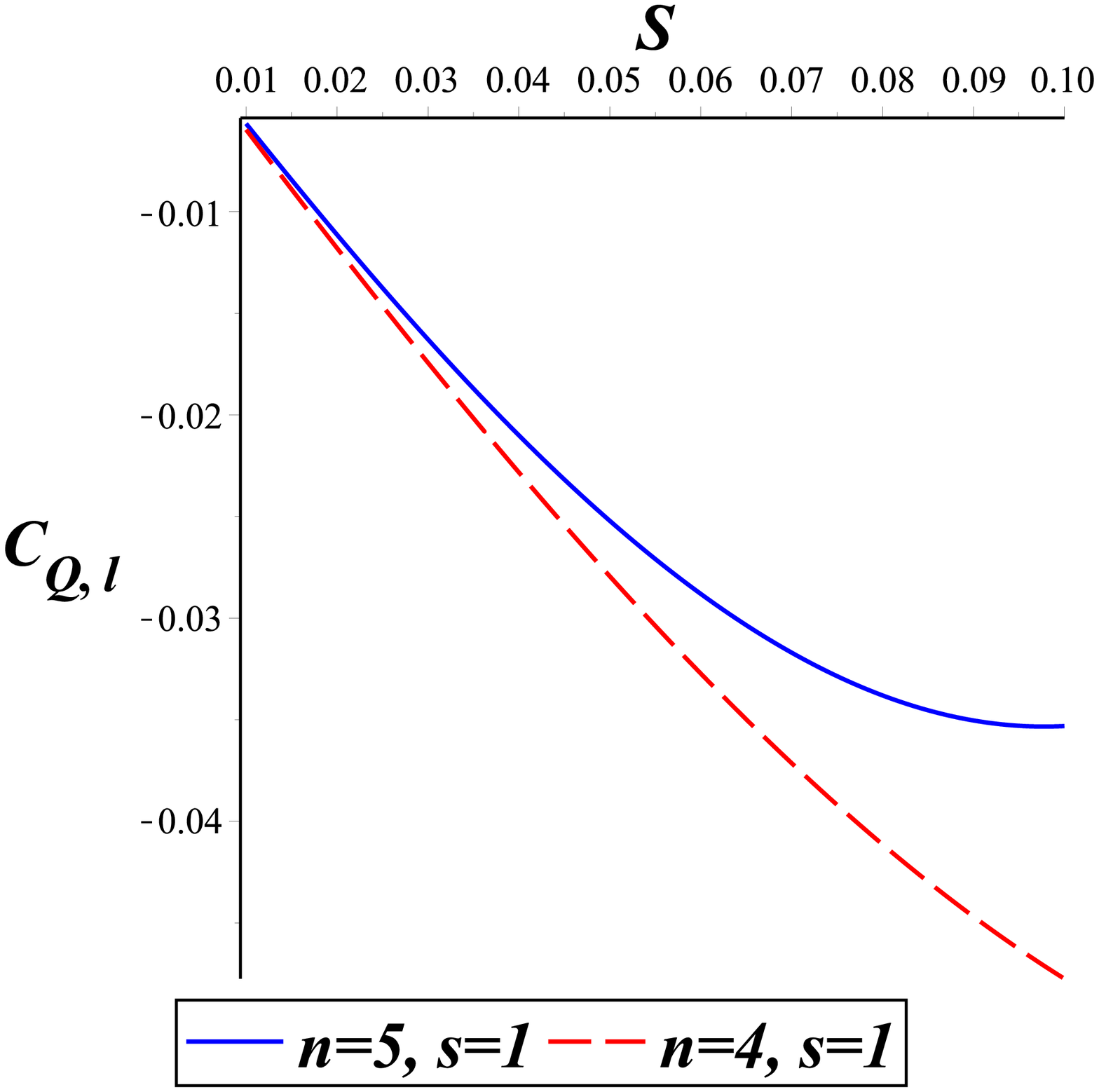}\includegraphics[scale=0.25]{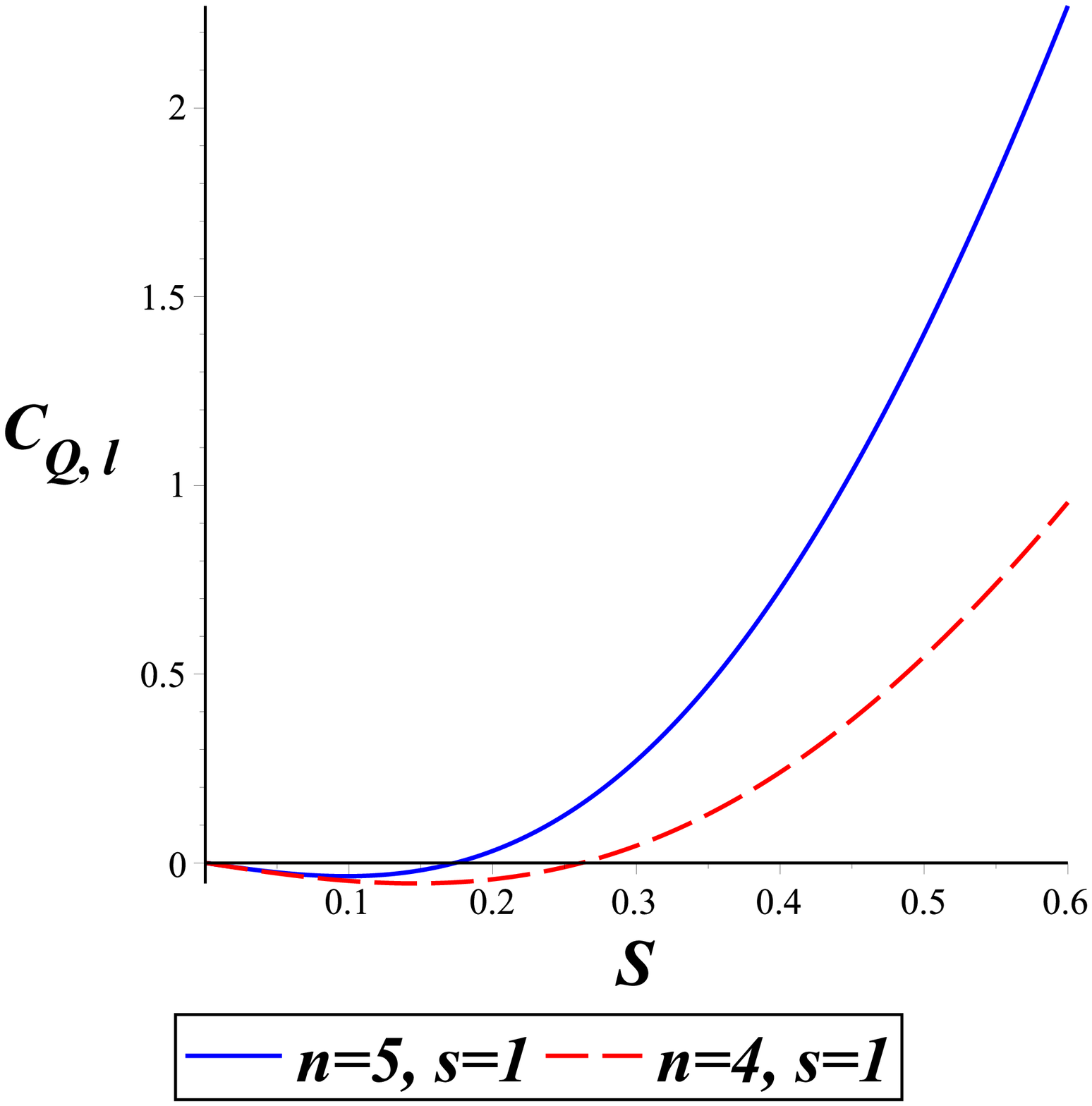}
\caption{{\small Heat Capacity $C_{Q\,,l}$ as a function of
entropy $S$, where we have considered $n=4$, $l=Q=1$ and $s=1$.}}}
\end{center}\end{figure}

On the other hand, figure 7 shows the behavior of the heat
capacity for $s=1$, which corresponds to  a black hole with linear
electromagnetic source.
As we can see, in the linear electromagnetic case there are no
phase transitions. This shows that the nonlinearity of the
electromagnetic source generate phase transitions.

%%%%%%%%%%%%%%%%%%%%%%%%%%%%%%%%%%%%%%%%%%%%%%%%%%%%%%%%%%%%%%%%%%%%%%%%%%%%%%%%%%%%%%%%%%%%%%%%%
%%%%%%%%%%%%%%%%%%%%%%%%%%%%%%%%%% SECTION III %%%%%%%%%%%%%%%%%%%%%%%%%%%%%%%%%%%%%%%%%%%%%%%%%%%
%%%%%%%%%%%%%%%%%%%%%%%%%%%%%%%%%%%%%%%%%%%%%%%%%%%%%%%%%%%%%%%%%%%%%%%%%%%%%%%%%%%%%%%%%%%%%%%%%

\section{geometrothermodynamics formalism} \label{GTD}

In brief, GTD is a formalism based on a space which is a $(2n +
1)-$ dimensional Riemannian contact manifold $(\mathcal{T} ,
\Theta, G )$, where $\mathcal{T}$ represets a differential
manifold, $\Theta$ is a contact form, i.e., $\Theta \wedge
(d\Theta)^n \neq 0$, and $G$ a Riemannian metric. If we introduce
in $\mathcal{T}$ the coordinates $Z^A = \{\Phi,E^a,I^a\}$ with $a
= 1, . . . n$ and $A = 0, . . . , 2n$, according to Darboux
theorem, the contact form $\Theta$ can be expressed as $\Theta =
d\Phi-\delta_{ab} I^a dE^b$. On the other hand, the main
ingredient of GTD is the Legendre invariance, which is considered
through the metric $G$ demanding that it must be invariant with
respect to Legendre transformations \cite{Arnold}. In particular,
three metrics have been found that are Legendre invariant
\cite{Quevedo5}. One of them is used to describe the
thermodynamics of black holes and can be written as

\bea \label{gtd6} G=\Theta^2+(\delta_{ab}E^a I^b)(\eta_{cd}dE^c
dI^d)\,,\eea where $\delta_{ab}={\mathrm{diag}}(1,1,\dots,1)$ and
$\eta_{ab}={\mathrm{diag}}(-1,1,\dots,1)$.
Using the metric
(\ref{gtd6}) GTD induces a Legendre invariant metric $g$ on an
$n-$ dimensional submanifold $\mathcal{E} \subset \mathcal{T}$ by
mean of the pullback $\varphi^{*} (G) =g$, which is associated with
the smooth embedding map $\varphi:\mathcal{E} \longrightarrow
\mathcal{T}$ and fulfills the condition $\varphi^{*} (\Theta)=0$.
If we choose the set $E^a $ as coordinates of $\mathcal{E}$, then
the embedding reads $\varphi : \{E^a\} \longrightarrow \{
\Phi(E^a), E^a, I^a(E^a)\}$ so that $\Phi(E^a)$ is the fundamental
equation and the induced metric becomes

\bea \label{gmetric} g=\beta_\Phi \Phi \eta_a^b \Phi_{,bc} dE^a
dE^b\,,\eea with $\Phi_{,a}=\frac{\partial \Phi}{\partial E^a}$,
$\beta_\Phi$ is a constant and we have used the Euler identity in
the form $\beta_a E^a \Phi_{,a}=\beta_\Phi \Phi$ for generalized
homogeneous function \cite{Quevedo5}.

We now apply the above formalism to the case of black holes with
PMI source. Let us consider the fundamental equation (\ref{mass2})
which, according to the analysis presented above, is a generalized
homogeneous function of degree 1 that does not involve a
redefinition of the thermodynamic variables, affecting the
physical properties of the thermodynamic system \cite{Quevedo5}.
The thermodynamic metric (\ref{gmetric}) is  3-dimensional and
reduces to

\bea \label{gmetric2} g&=&\frac{M
[4]^{\frac{n}{n-1}}\omega_{n-1}^{\frac{1}{n-1}}}{\pi(n-1)}\Bigg\{\Big[(n-2)
S^{-\frac{n}{n-1}}-\frac{n}{l^2}S^{-\frac{n-2}{n-1}}+\nonumber
\\ &+& \frac{(2s-n)(4s-2sn-1)\tilde{f}_n}{(2s-1)^2}S^{-\frac{6s-4ns-n-2}{(n-1)(2s-1)}}
Q^{\frac{2s}{2s-1}}\Big] dS^2 -
\frac{2s\tilde{f}_n}{(2s-1)^2}S^{\frac{2s-n}{(n-1)(2s-1)}}
Q^{-\frac{2(s-1)}{2s-1}}dQ^2+ \nonumber
\\ &+&\frac{6}{l^4}S^{\frac{n}{(n-1)}}dl^2\Bigg\}\,.\eea The curvature scalar corresponding to the metric (\ref{gmetric2}) takes the form,

\bea \label{scalar} R=\frac{N(S,Q,l)}{D_1 D_2}\,,\eea where,

\bea \label{denom123} D_1&=& \Big[l^2 (n-2) S^{-\frac{n}{(n-1)}}-n
S^{-\frac{(n-2)}{(n-1)}}+
\frac{(2s-n)(4s-2sn-1)}{(2s-1)^2}l^2\tilde{f}_n
S^{\frac{6s-4ns+n-2}{(n-1)(2s-1)}}Q^{\frac{2s}{2s-1}}\Big]^2\,,\\
D_2&=&l^2 S^{\frac{n-2}{n-1}}+S^{\frac{n}{n-1}}-l^2 \tilde{f}_n
S^{\frac{2s-n}{(n-1)(2s-1)}}Q^{\frac{2s}{2s-1}}\,,\eea and
$N(S,Q,l)$ is a function that is different from zero at the points
where denominator vanishes and cannot be written in a compact
form. There are two curvature singularities in this case. The
first one occurs if $D_2=0$ and corresponds to $M = 0$, as follows
from equation (\ref{mass2}), see also figure 4. This means that
this singularity is non physical since no black hole is present in
this case. A second singularity is located at the roots of the
equation $D_1=0$, it coincides with the points where $C_{Q,l}
\longrightarrow \infty$, i. e., with the points where second order
phase transitions take place. The general behavior of the
curvature scalar is illustrated in figures 9 y 10.

\begin{figure}[h]\begin{center}{\includegraphics[scale=0.25]{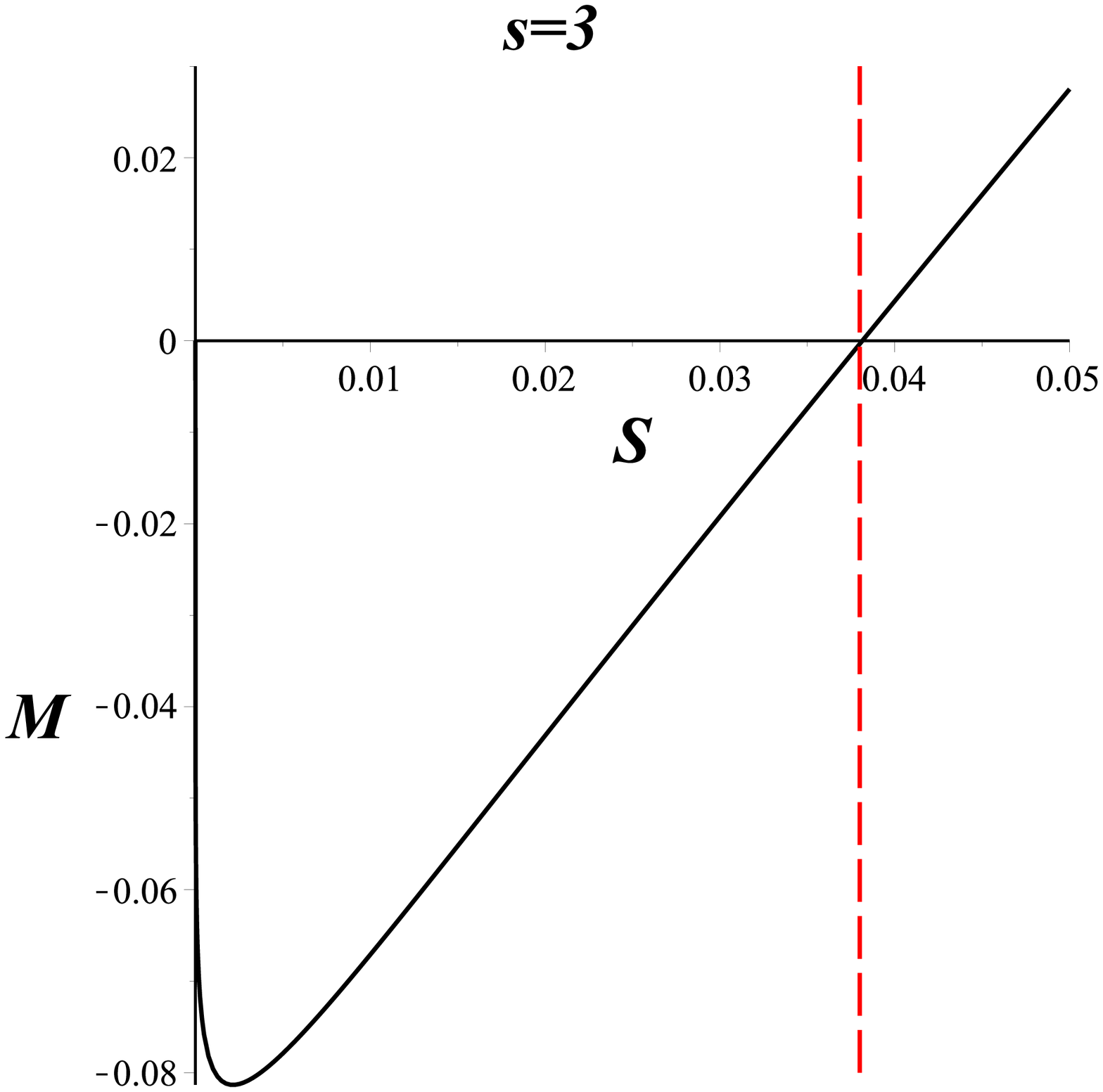}\includegraphics[scale=0.25]{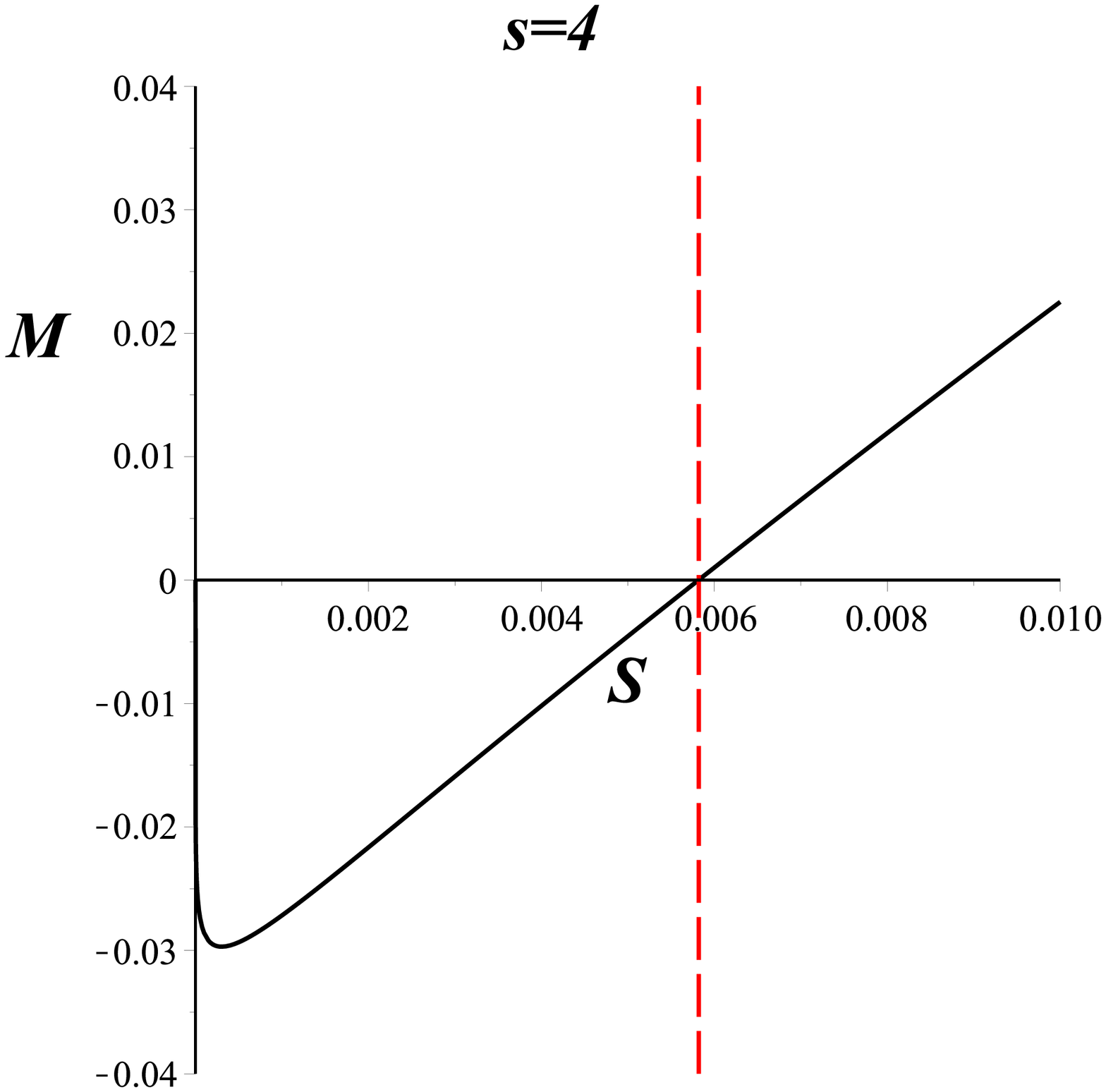}
\caption{{\small The mass $M$ as a function of the entropy $S$. We
have considered $n=4$, $s=3$(left) and $n=5$, $s=4$(right), in
both cases $l=Q=1$ and .}}}
\end{center}\end{figure}

\begin{figure}[h]\begin{center}{\includegraphics[scale=0.25]{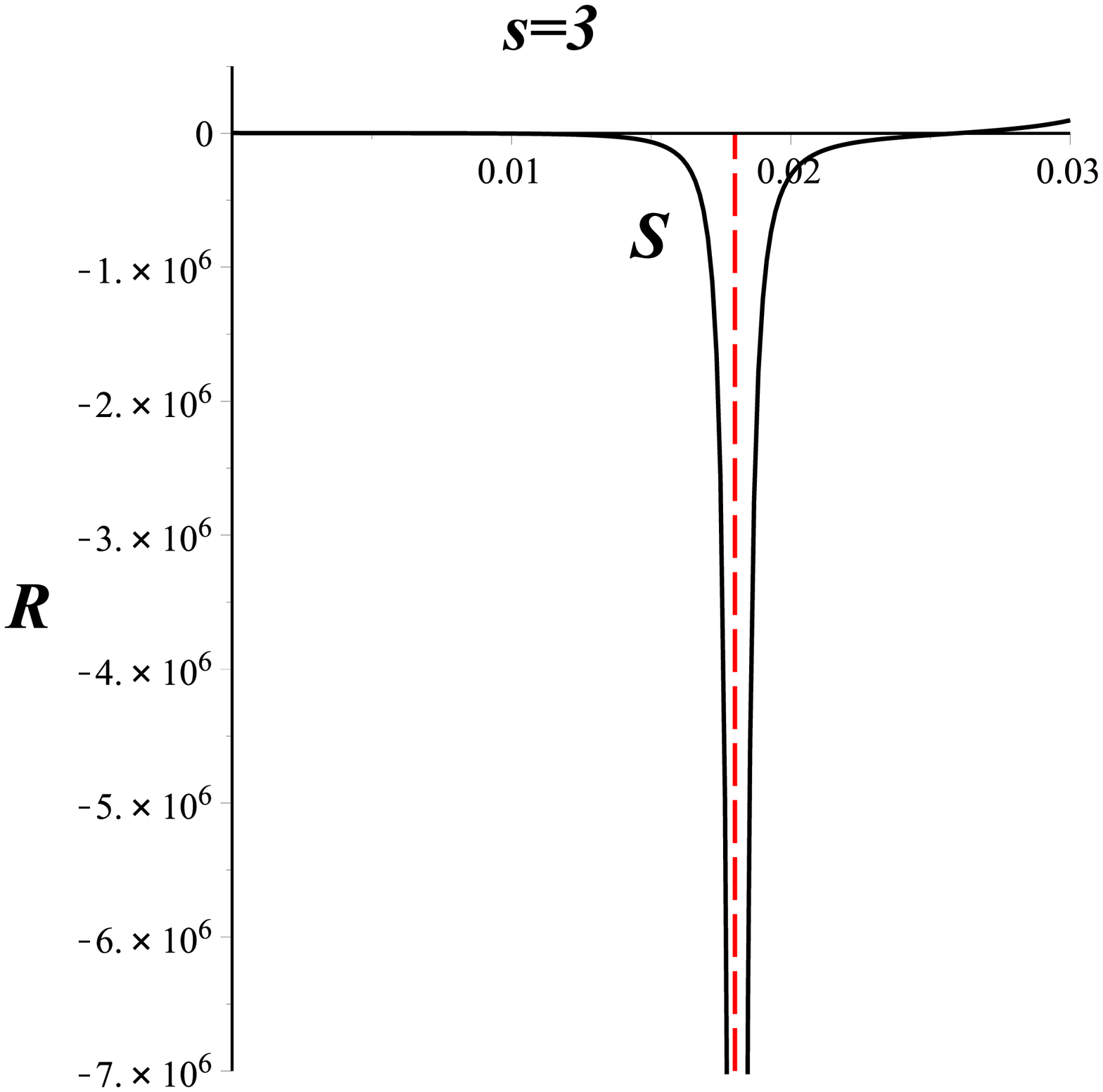}
\includegraphics[scale=0.25]{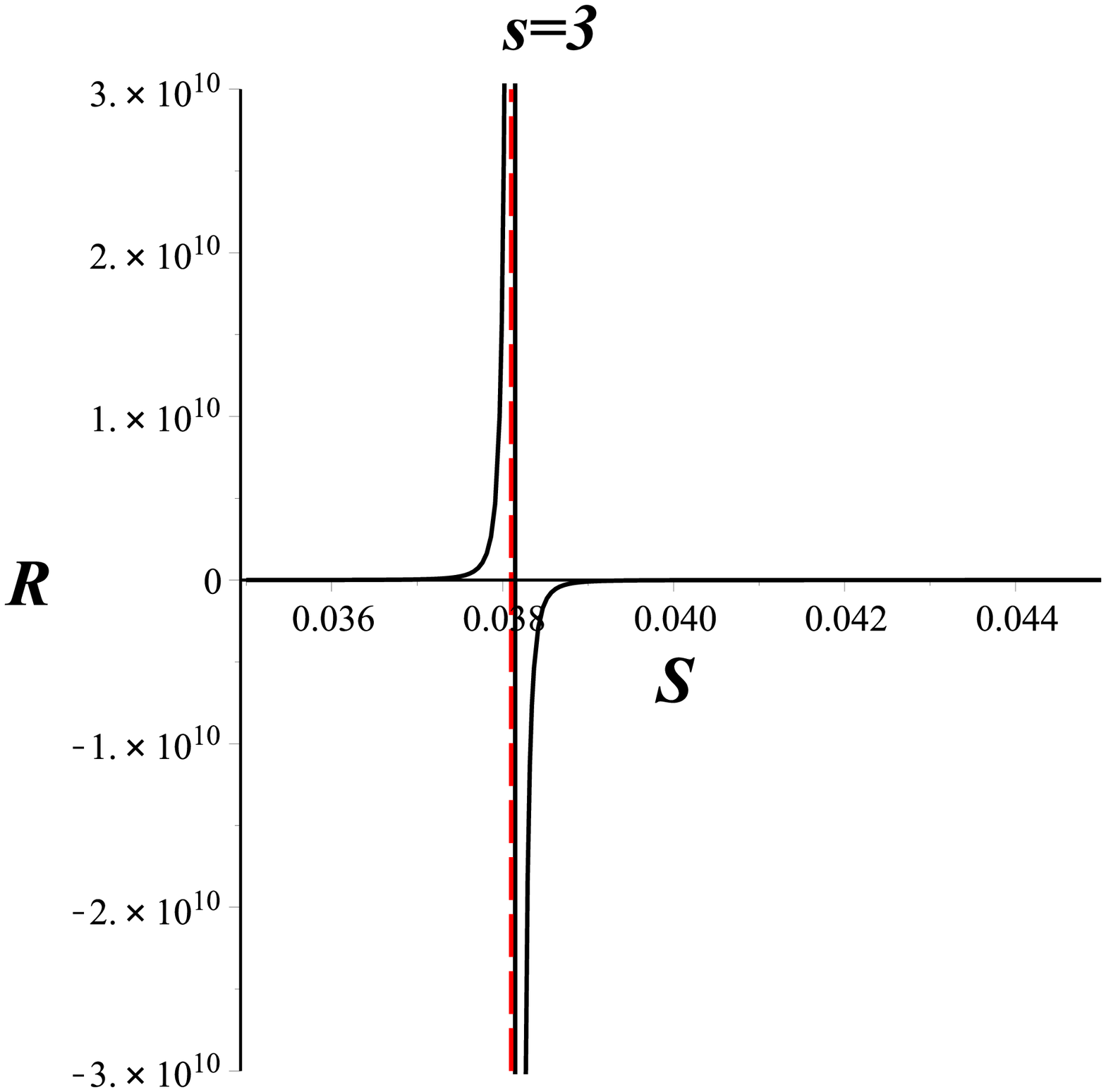}\includegraphics[scale=0.25]{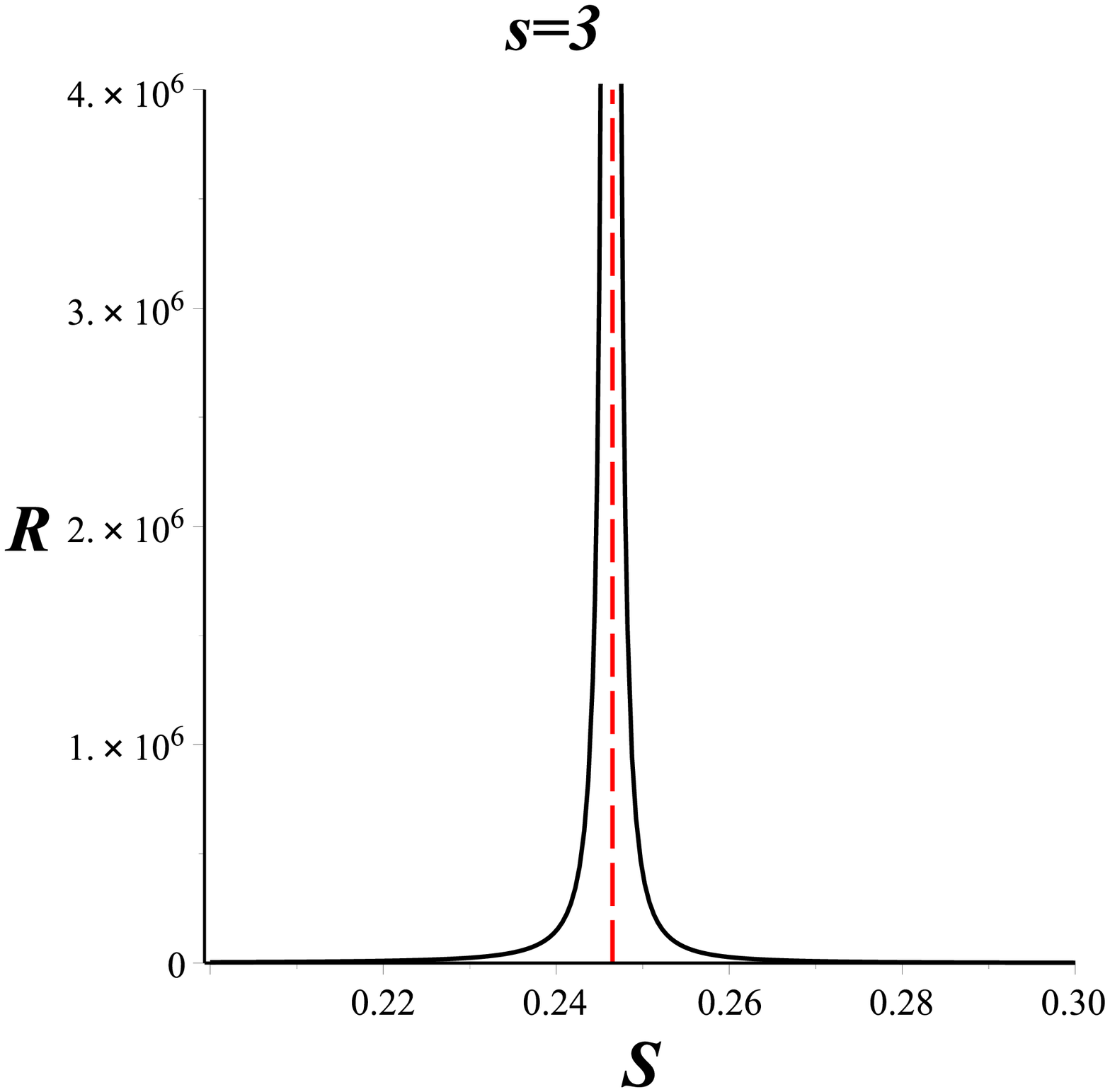}
\caption{{\small Behavior of the curvature scalar $R$, as a
function of entropy $S$, for different intervals. We have
considered $n=4$, $l=Q=1$ and $s=3$.}}}
\end{center}\end{figure}

\newpage

\begin{figure}[h]\begin{center}{\includegraphics[scale=0.25]{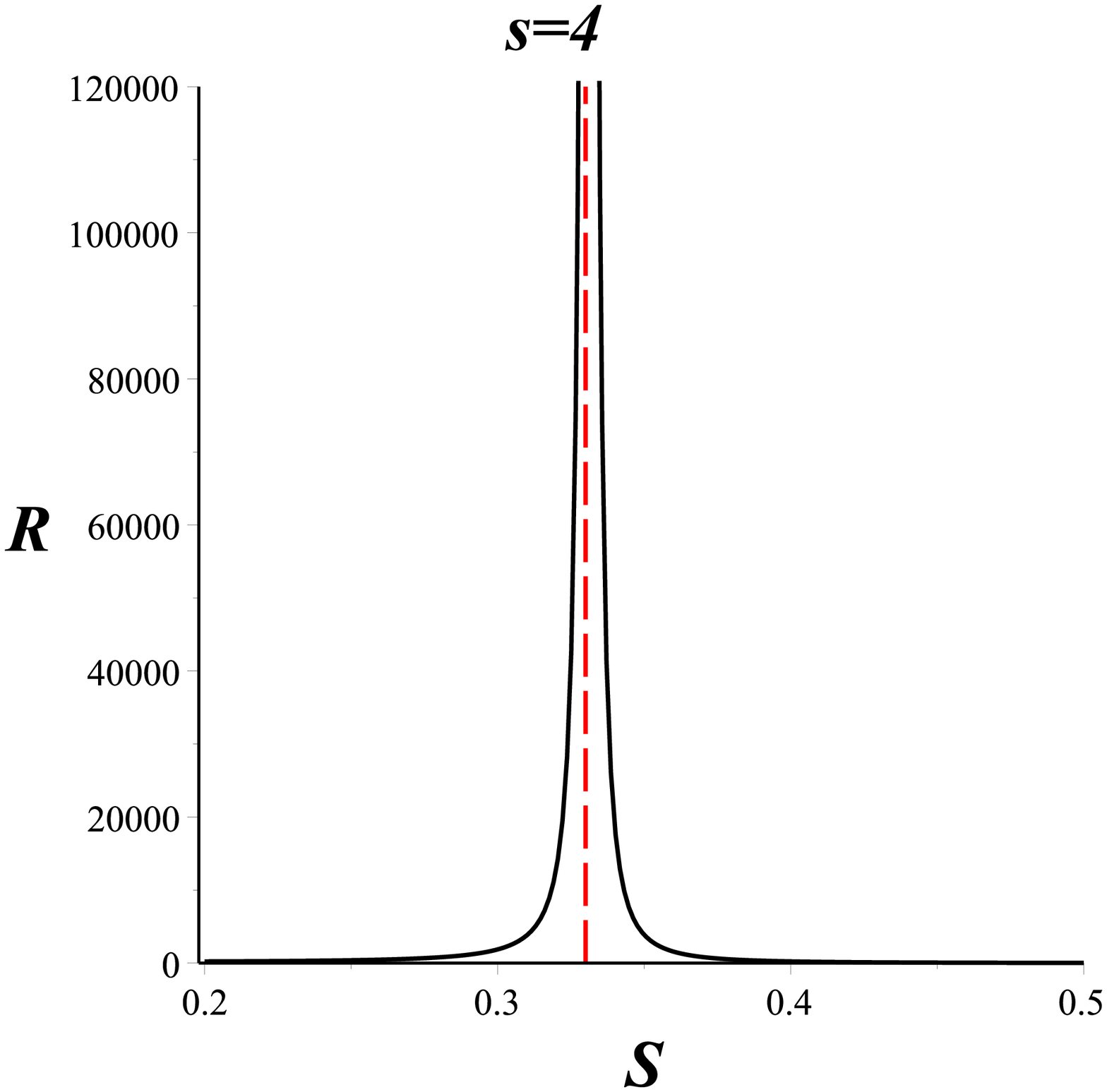}
\includegraphics[scale=0.25]{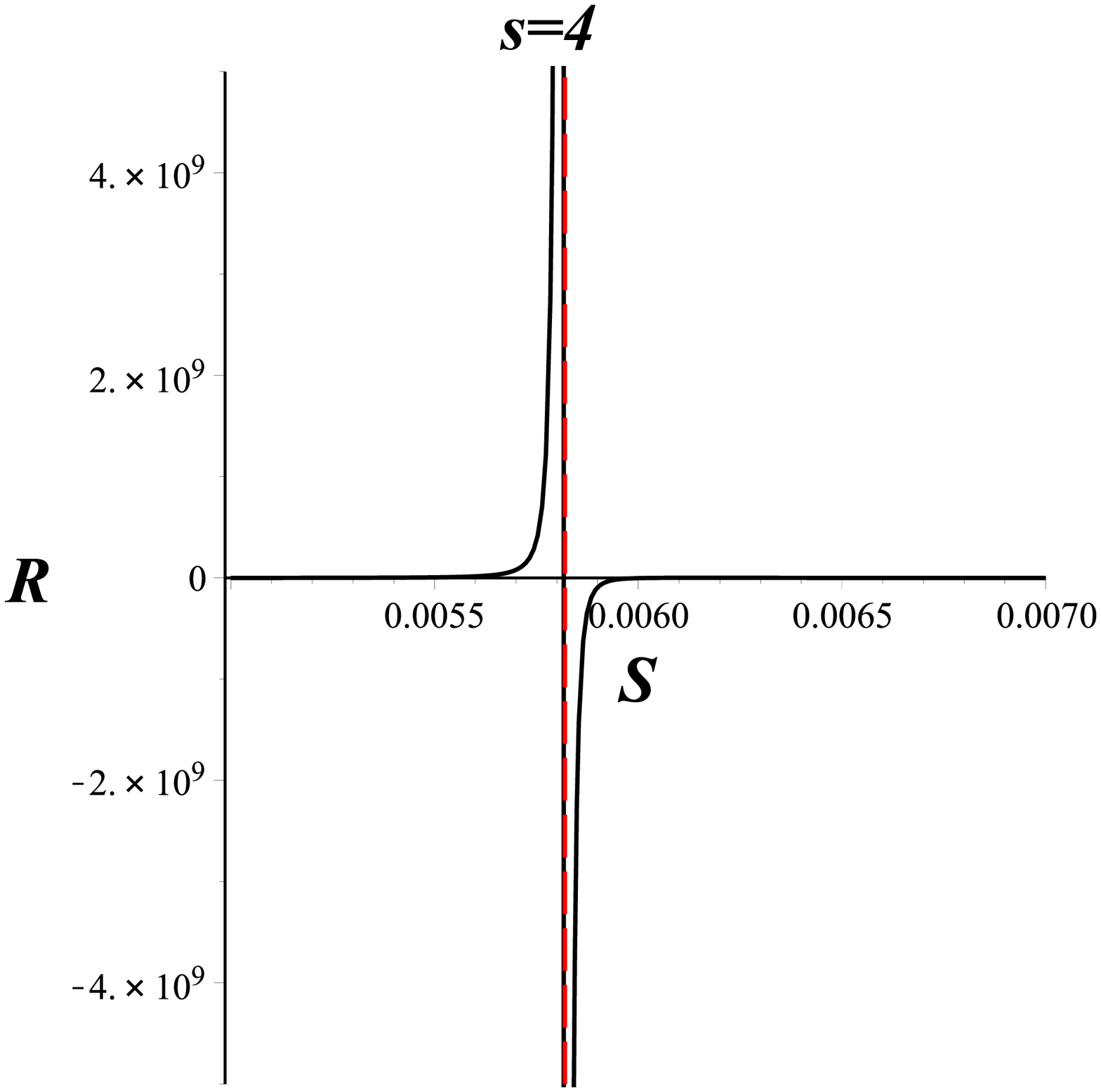}\includegraphics[scale=0.25]{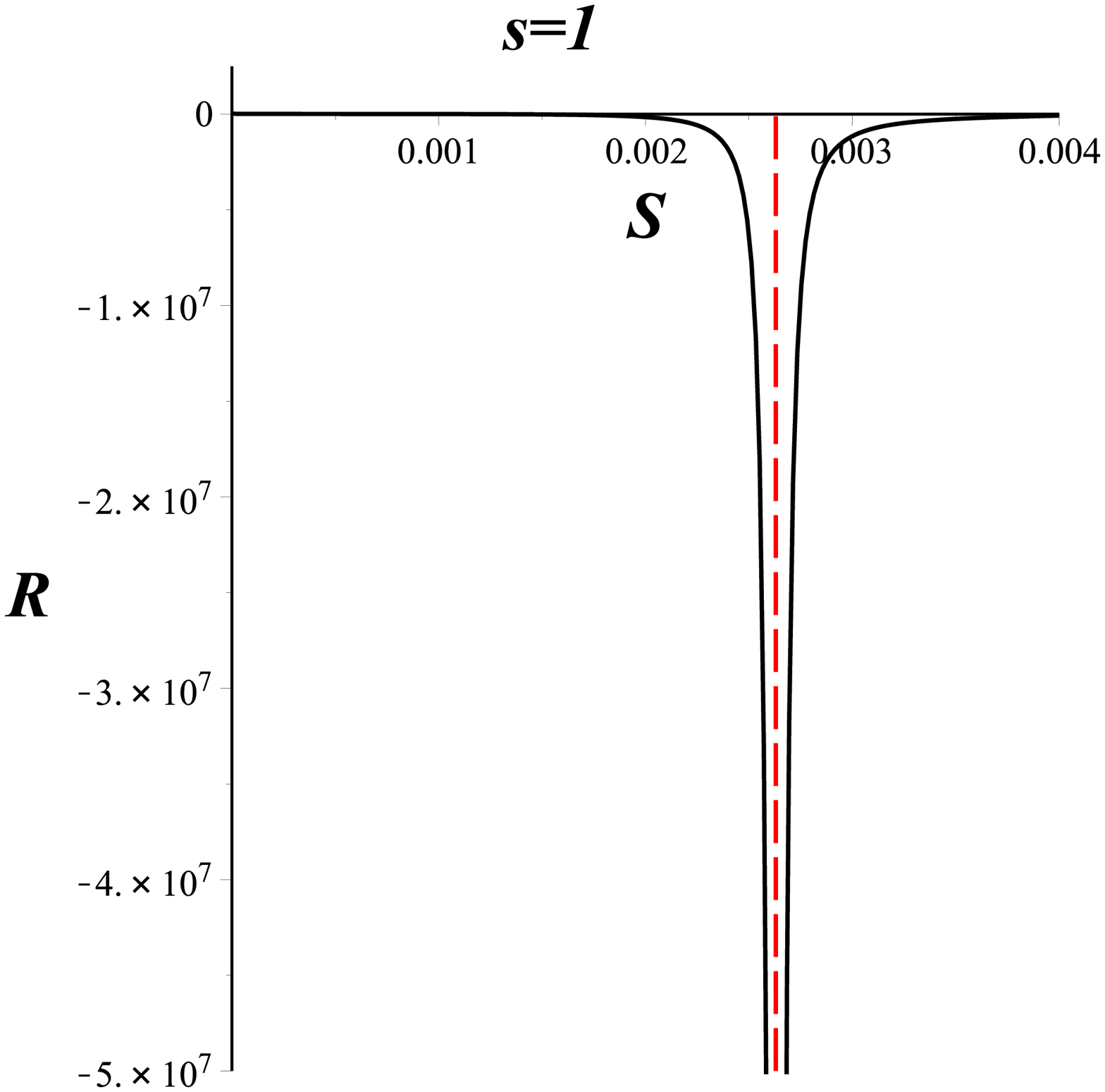}
\caption{{\small Behavior of the curvature scalar $R$, as a
function of entropy $S$, for different intervals. We have
considered $n=5$, $l=Q=1$ and $s=4$.}}}
\end{center}\end{figure}

According to GTD, these results show that there exist curvature
singularities at those points where second order phase transitions
occur, because the denominators of the heat capacity and the
curvature scalar coincide \cite{quevedo1}.

\begin{figure}[h]\begin{center}{\includegraphics[scale=0.25]{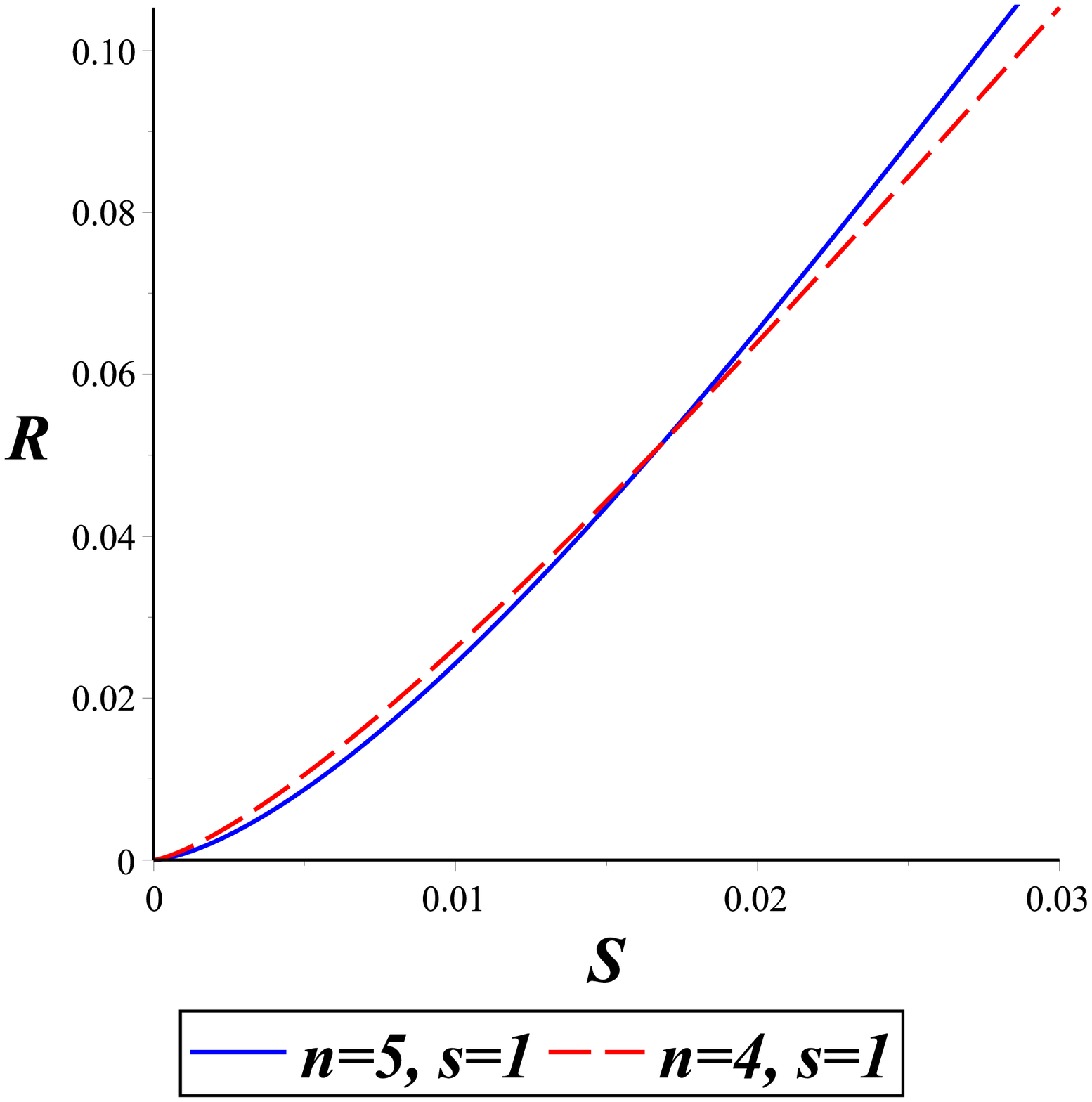}
\includegraphics[scale=0.25]{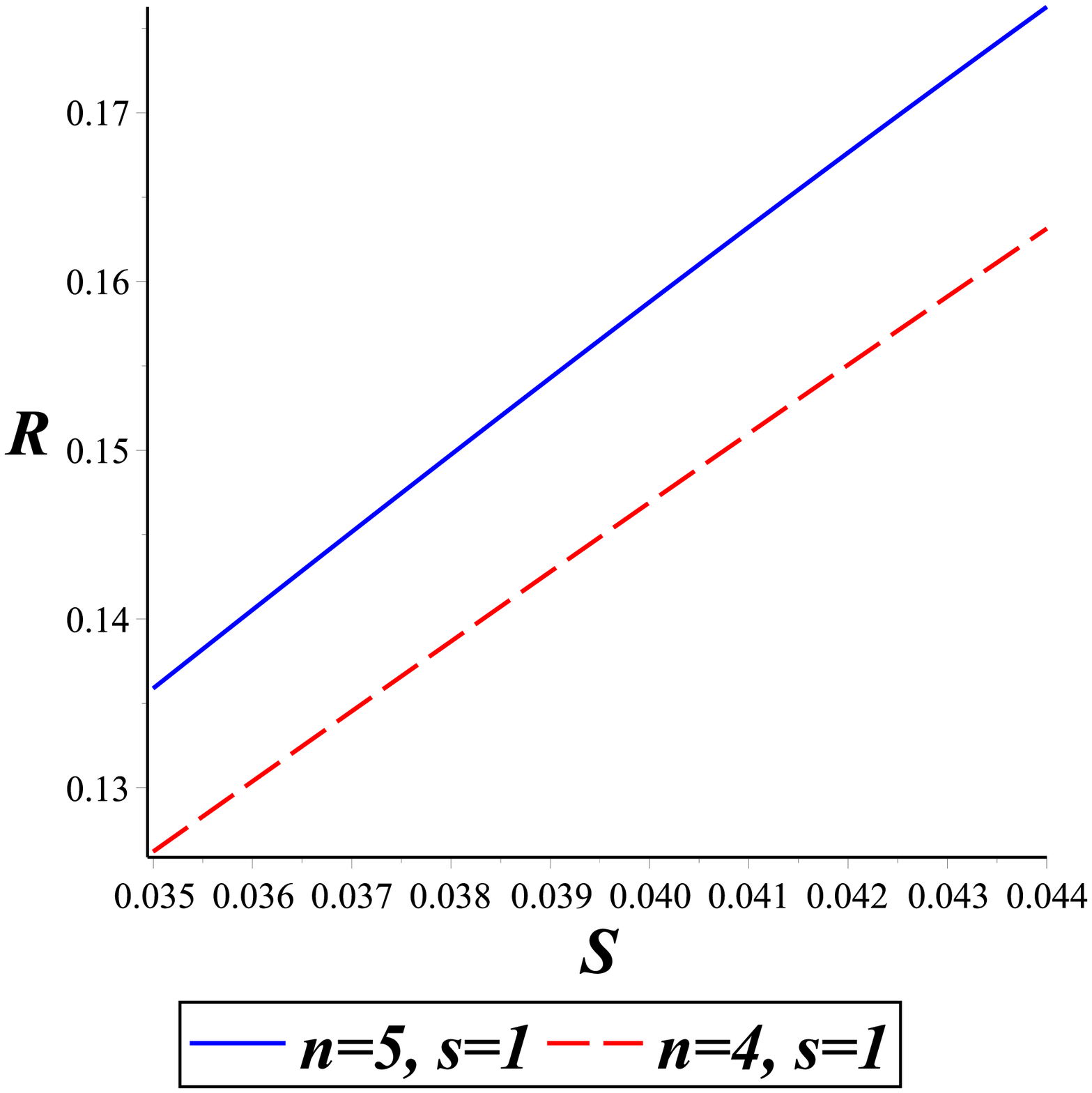}\includegraphics[scale=0.25]{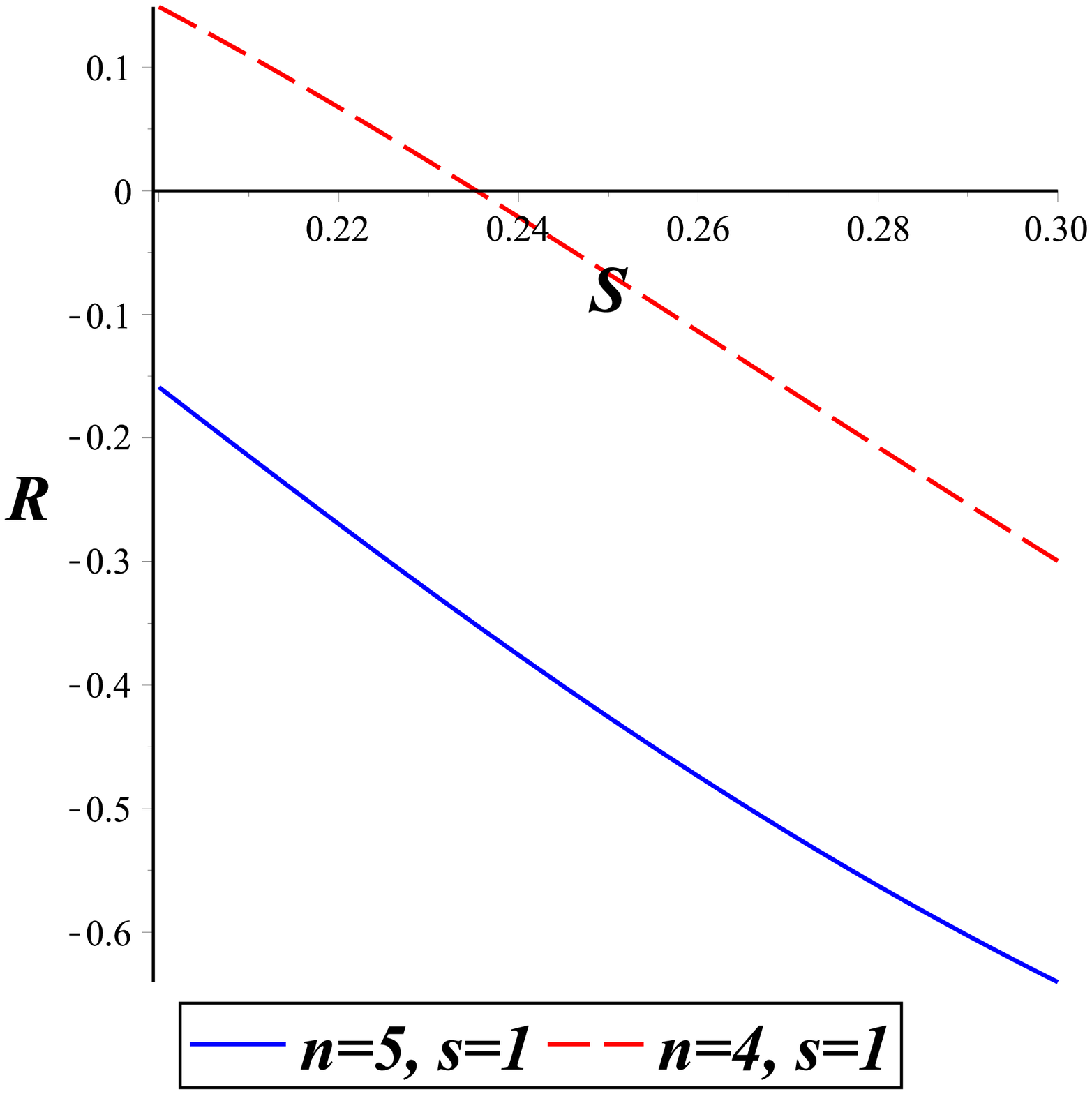}
\caption{{\small Behavior of the curvature scalar $R$, as a
function of entropy $S$, for different intervals. We have
considered $n=4$, $l=Q=1$ and $s=1$.}}}
\end{center}\end{figure}

Figure 11 shows the behavior of the curvature scalar for a black
hole with linear electromagnetic source ($s=1$). As we can see in the linear
electromagnetic case there are no singularities. Therefore, GTD
also reproduces correctly this case.

%%%%%%%%%%%%%%%%%%%%%%%%%%%%%%%%%%%%%%%%%%%%%%%%%%%%%%%%%%%%%%%%%%%%%%%%%%%%%%%%%%%%%%%%%%%%%%%%%
%%%%%%%%%%%%%%%%%%%%%%%%%%%%%%%%%% SECTION V %%%%%%%%%%%%%%%%%%%%%%%%%%%%%%%%%%%%%%%%%%%%%%%%%%%
%%%%%%%%%%%%%%%%%%%%%%%%%%%%%%%%%%%%%%%%%%%%%%%%%%%%%%%%%%%%%%%%%%%%%%%%%%%%%%%%%%%%%%%%%%%%%%%%%

\section{Conclusions}
\label{conclusions}

In this paper, we investigated the thermodynamics and
geometrothermodynamics of a spherically symmetric AdS black hole
with a PMI source. We analyzed the fundamental equation that
relates the total mass, the entropy and charge. We showed that for
the fundamental equation to be a generalized homogeneous function
\cite{Quevedo5}, it is necessary to consider the variable $l$,
related to the cosmological constant, as an extensive
thermodynamic variable, in order for the fundamental equation to
depend on extensive variables only. As a result, we obtained a
fundamental equation whose mathematical properties resemble those
of classical thermodynamic systems. Considering the curvature
radius $l$ as an extensive thermodynamic variable implies that the
equilibrium space must be extended by one dimension. A similar
result was obtained recently \cite{Dolan}, by assuming that the
energy of a black hole is not represented by its total mass, but
by the corresponding enthalpy, indicating that the cosmological
constant is an intensive thermodynamic variable similar to the
pressure. Our results corroborate from a more formal mathematical
point of view the intuitive analysis performed in \cite{Dolan}.

We investigated the properties of the extended 3-dimensional
equilibrium space in the framework of GTD and we showed that in
the space of equilibrium states of a black hole with PMI source,
there exists a thermodynamic metric whose curvature turns out to
be nonzero, indicating the presence of thermodynamic interaction.
We also found that the curvature is singular at those points where
phase transitions of the heat capacity occur. This has been shown
by considering a particular metric in the thermodynamic phase
space, and applying the formalism of geometrothermodynamics. An
important property of our choice of thermodynamic metric is that
it is invariant with respect to Legendre transformations so that
the properties of our geometric description of thermodynamics are
independent of the choice of thermodynamic potential and
representation.

We conclude that the thermodynamic properties of this particular
class of a black holes with no linear source are correctly
described within the GTD formalism.

\section*{Acknowledgements}

This work was partially supported by Conacyt-Mexico, Grant No.
A1-S-31269, and by UNAM-DGAPA-PAPIIT, Grant No. 114520.


\begin{thebibliography}{99}

%*******************************************************************************************************

\bibitem{Ayon2} E. Ayon-Beato and A. Garcia, Phys. Lett. B 464, 25 (1999).

\bibitem{Bronnikov} K. A. Bronnikov, Phys. Rev. D63, 044005 (2001)

\bibitem{Hassaine2} M. Hassaine and C. Martinez, Class. Quantum Gravit. 25, 195023
(2008)

\bibitem{Hassaine} M. Hassaine and C. Martinez, Phys.\ Rev.\ D {\bf 75}, 027502 (2007).

\bibitem{Hendi0} S. H. Hendi and H. R. Rastegar-Sedehi, Gen.\ Rel.\ Grav.\  {\bf 41}, 1355
(2009).

\bibitem{Maeda} H. Maeda, M.~ Hassaine and C. Martinez, JHEP {\bf 1008}, 123 (2010).

\bibitem{Kats} Y. Kats, L. Motl and M. Padi, JHEP 0712, 068 (2007);

\bibitem{Maldacena} J. M. Maldacena, AIP Conf. Proc.  {\bf 484}, 51 (1999) [Int. J. Theor. Phys.  {\bf 38}, 1113 (1999)].

\bibitem{Hawking} S. W. Hawking and D. N. Page, Commun.\ Math.\ Phys.\  {\bf 87}, 577 (1983).


\bibitem{Chamblin1} A. Chamblin, R. Emparan, C. V. Johnson and R. C.~ Myers, Phys. Rev. D {\bf 60}, 064018 (1999).

\bibitem{Kastor} D. Kastor, S. Ray and J. Traschen,  Class. Quant. Grav.  {\bf 26}, 195011 (2009).

\bibitem{Dolan1} B. P. Dolan, Class.  Quant.  Grav.   {\bf 28}, 125020 (2011).

\bibitem{davies} P. C. W. Davies, Proc.  Roy.  Soc.  Lond.  A {\bf 353}, 499 (1977).

\bibitem{Weinhold} F. Weinhold, J. Chem. Phys. 63, 2479, 2484, 2488, 2496 (1975); 65, 558 (1976).

\bibitem{Ruppeiner} G. Ruppeiner, Phys. Rev. A 20, 1608 (1979).

\bibitem{quevedo2} H. Quevedo, J.  Math.  Phys.  {\bf 48}, 013506 (2007).

\bibitem{Hendi} S. H. Hendi and M. H. Vahidinia, Phys.  Rev.  D {\bf 88}, 084045 (2013).

\bibitem{Quevedo5} H. Quevedo, M. N. Quevedo, and A. Sánchez, Eur. Phys. J. C 77, 158 (2017).

\bibitem{Quevedo6} H. Quevedo, M. N. Quevedo, and A. S\'anchez,Eur. Phys. J. C (2019)
79:229.

\bibitem{callen} H.B. Callen,{\em Thermodinamics} (John Wiley and Sons, Inc., New York, 1981), pag. 132.

\bibitem{Arnold} V. I. Arnold, Mathematical Methods of Classical Mechanics (Springer Verlag, New York, 1980).

\bibitem{quevedo1} H. Quevedo, A. Sanchez, S. Taj and A. Vazquez, Gen.  Rel.  Grav.   {\bf 43}, 1153 (2011).

\bibitem{Dolan} B. P. Dolan, Where is the PdV term in the first law of black hole thermodynamics?, arXiv:1209.1272
[gr-qc](2016).


%*******************************************************************************************************




\end{thebibliography}
\end{document}